\documentclass[a4paper,12pt]{article}
\pdfoutput=1
\usepackage{feynmp-auto,expdlist}
\usepackage{amsmath, amsfonts, amssymb}
\usepackage{graphicx}
\usepackage{enumerate,cancel}
\usepackage{hyperref}
\usepackage{latexsym}
\usepackage{dsfont}
\usepackage{hepnicenames}
\usepackage{enumerate}
\usepackage{soul}
\usepackage[normalem]{ulem}
\usepackage{comment}

\usepackage{units}

\usepackage{soul}

\newcommand{\fNL}{f_{\rm NL}}

\usepackage{mathrsfs,graphicx,rotating,amsmath,amsfonts,mathtools,booktabs,amssymb,wasysym}
\usepackage{hyperref}\usepackage{slashed}
\usepackage[nosort]{cite}
\usepackage[table,xcdraw,s]{xcolor}
\usepackage{bm}
\usepackage{graphicx}
\usepackage{multirow,multicol}
\hypersetup{colorlinks,bookmarksopen,bookmarksnumbered,
	linkcolor=blus,pdfstartview=FitH,urlcolor=rossos,citecolor=verde}
\allowdisplaybreaks

\renewcommand{\[}{\left[}

\newcommand{\Lag}{\mathscr{L}}

\newcommand{\mio}[1]{}

\newcommand{\med}[1]{\langle #1\rangle}

\newcommand{\bpm}{\begin{pmatrix}}
\newcommand{\epm}{\end{pmatrix}}

\usepackage{mathrsfs}

\newcommand{\fig}[1]{~\ref{fig:#1}}
\newcommand{\sfrac}[2]{#1/#2}

\allowdisplaybreaks
\usepackage{multicol}
\usepackage{color}
\definecolor{rosso}{cmyk}{0,1,1,0.4}
\definecolor{rossos}{cmyk}{0,1,1,0.55}
\definecolor{rossoc}{cmyk}{0,1,1,0.2}
\definecolor{blu}{cmyk}{1,1,0,0.3}
\definecolor{blus}{cmyk}{1,1,0,0.6}
\definecolor{bluc}{cmyk}{1,1,0,0.1}
\definecolor{verde}{cmyk}{0.92,0,0.59,0.25}
\definecolor{verdec}{cmyk}{0.92,0,0.59,0.15}
\definecolor{verdes}{cmyk}{0.92,0,0.59,0.4}

\newcommand{\bp}{\bar{M}_{\rm Pl}}
\newcommand{\riga}[1]{\noalign{\hbox{\parbox{\textwidth}{#1}}}\nonumber}

\newcommand{\eq}[1]{~{\rm (\ref{eq:#1})}}

\newcommand{\GeV}{\,{\rm GeV}}
\newcommand{\TeV}{\,{\rm TeV}}

\newcommand{\beq}{\begin{equation}}
\newcommand{\eeq}{\end{equation}}
\newcommand{\mb}[1]{\mbox{\boldmath $#1$}}
\newcommand{\bea}{\begin{eqnarray}}
\newcommand{\eea}{\end{eqnarray}}
\newcommand{\be}{\begin{equation}}
\newcommand{\ee}{\end{equation}}
\font\tenrsfs=rsfs10 at 12pt
\font\sevenrsfs=rsfs7
\font\fiversfs=rsfs5
\newfam\rsfsfam
\textfont\rsfsfam=\tenrsfs
\scriptfont\rsfsfam=\sevenrsfs
\scriptscriptfont\rsfsfam=\fiversfs
\def\be#1\ee{\begin{equation}#1\end{equation}}
\def\bl#1\el{\begin{align}#1\end{align}}
\def\ba#1\ea{\begin{align*}#1\end{align*}}

\renewenvironment{thebibliography}[1]
{\begin{multicols}{2}[\section*{\refname}]%
		\@mkboth{\MakeUppercase\refname}{\MakeUppercase\refname}%
		\list{\@biblabel{\@arabic\c@enumiv}}%
		{\settowidth\labelwidth{\@biblabel{#1}}%
			\leftmargin\labelwidth
			\advance\leftmargin\labelsep
			\@openbib@code
			\usecounter{enumiv}%
			\let\p@enumiv\@empty
			\renewcommand\theenumiv{\@arabic\c@enumiv}}%
		\sloppy
		\clubpenalty4000
		\@clubpenalty \clubpenalty
		\widowpenalty4000%
		\sfcode`\.\@m}
	{\def\@noitemerr
		{\@latex@warning{Empty `thebibliography' environment}}%
		\endlist\end{multicols}}

\makeatletter
\font\ital=cmu10

\def\hhref#1{\href{http://arxiv.org/abs/#1}{arXiv:#1}}
\usepackage{xstring}
\newcommand{\hhrefq}[1]{\IfSubStr{#1}{:}{\href{http://inspirehep.net/search?ln=en&ln=en&p=#1&of=hb&action_search=Search&sf=&so=d&rm=&rg=25&sc=0}{InSpire:#1}}{\hhref{#1}}}

\def\art{\@ifnextchar[{\eart}{\oart}}
\def\eart[#1]#2#3#4#5#6{{\rm #2}, {\em #3 \bf #4} {\rm (#6) #5} ({\em #1})}
\def\article{\@ifnextchar[{\earticle}{\oarticle}}
\def\oarticle#1#2#3#4#5#6{{\rm #1}, {\ital `#6'}, {\rm #2 #3 (#5) #4}}
\def\earticle[#1]#2#3#4#5#6#7{{\rm #2}, {\ital `#7'}, {\rm #3 #4 (#6) #5}  [\hhrefq{#1}]}
\def\hepart[#1]#2{{\rm #2, \sl#1}}
\def\heparticle[#1]#2#3{#2, {\ital `#3'} [\hhrefq{#1}]}
\newcommand{\doi}[1]{\href{http://dx.doi.org/#1}{[link]}}

\newcommand{\hhrefqq}[1]{\IfBeginWith{#1}{10.}{\href{https://doi.org/#1}{doi:#1}}{\hhrefq{#1}}}

\renewenvironment{thebibliography}[1]
{\begin{multicols}{2}[\section*{\refname}]%
		\@mkboth{\MakeUppercase\refname}{\MakeUppercase\refname}%
		\list{\@biblabel{\@arabic\c@enumiv}}%
		{\settowidth\labelwidth{\@biblabel{#1}}%
			\leftmargin\labelwidth
			\advance\leftmargin\labelsep
			\@openbib@code
			\usecounter{enumiv}%
			\let\p@enumiv\@empty
			\renewcommand\theenumiv{\@arabic\c@enumiv}}%
		\sloppy
		\clubpenalty4000
		\@clubpenalty \clubpenalty
		\widowpenalty4000%
		\sfcode`\.\@m}
	{\renewcommand{\@noitemerr}
		{\@latex@warning{Empty `thebibliography' environment}}%
		\endlist\end{multicols}}

\newcounter{alphaequation}[equation]
\def\thealphaequation{\theequation\alph{alphaequation}}
\renewcommand{\thealphaequation}{\theequation\hbox to
	0.6em{\hfil\alph{alphaequation}\hfil}}
\newcommand{\eqnsystem}[1]{
	\renewcommand{\@eqnnum}{{\rm (\thealphaequation)}}
	\renewcommand{\@@eqncr}{\let\@tempa\relax \ifcase\@eqcnt \def\@tempa{& & &} \or
		\newcommand{\@tempa}{& &}\or \newcommand{\@tempa}{&}\fi\@tempa
		\if@eqnsw\@eqnnum\refstepcounter{alphaequation}\fi
		\global\@eqnswtrue\global\@eqcnt=0\cr}
	\refstepcounter{equation} \let\@currentlabel\theequation \def\@tempb{#1}
	\ifx\@tempb\empty\else\label{#1}\fi
	\refstepcounter{alphaequation}
	\let\@currentlabel\thealphaequation
	\global\@eqnswtrue\global\@eqcnt=0 \tabskip\@centering\let\\=\@eqncr
	$$\halign to \displaywidth\bgroup \@eqnsel\hskip\@centering
	$\displaystyle\tabskip\z@{##}$&\global\@eqcnt\@ne
	\hskip2\arraycolsep\hfil${##}$\hfil& \global\@eqcnt\tw@\hskip2\arraycolsep
	$\displaystyle\tabskip\z@{##}$\hfil
	\tabskip\@centering&\llap{##}\tabskip\z@\cr}
\def\endeqnsystem{\@@eqncr\egroup$$\global\@ignoretrue} \makeatother

\definecolor{Gray}{gray}{0.95}
\newcommand{\bbox}[1]{\fcolorbox{gray}{Gray}{~$\displaystyle #1$~}}

\newcommand{\ag}[1]{\textcolor{orange}{[Anish: #1]}}

\usepackage{tocloft}
\begin{document}
\thispagestyle{empty}
\begin{center}  
CTPU-PTC-24-24\\[3ex]
{\LARGE\bf\color{rossos} Cosmological collider non-Gaussianity \\ from multiple scalars  and $R^2$ gravity} \\
\vspace{0.6cm}
{\bf Shuntaro Aoki}$^a$,
{\bf Anish Ghoshal}$^b$ and {\bf Alessandro Strumia}$^c$  \\[6mm]

{\it $^a$ Particle Theory and Cosmology Group, Center for Theoretical Physics of the Universe,
Institute for Basic Science, Daejeon, 34126, Korea\\[1ex]
$^b$ Institute of Theoretical Physics, Faculty of Physics, University of Warsaw, Poland\\[1ex]
$^c$ Dipartimento di Fisica, Universit\`a di Pisa, Pisa, Italia}

\vspace{0.5cm}
{\large\bf Abstract}
\begin{quote}\large
Cosmological collider signals of primordial non-Gaussianity arise at tree level  when an extra scalar has  Hubble mass during inflation. 
We critically review the formalism finding that a large class 
of inflationary theories, based on Planck-scale physics,
predict a scalar bi-spectrum around the gravitational floor level.
This mild signal arises for example in $R^2$ gravity,
in the regime where its gravitational scalar has Hubble-scale mass.
Signals much above the gravitational floor
 arise in theories where scalars undergo multiple turns during inflation,
thanks to sub-Planckian physics.
\end{quote}
\end{center}
\setcounter{page}{1}
\tableofcontents


\section{Introduction}
While the Large Hadron Collider with energy $\sqrt{s}\approx 13\TeV$ and integrated luminosity ${\cal L}\sim 1/{\rm ab}$
found no new physics around the weak scale and no next higher-energy collider is approved, 
cosmological observations offer a `cosmological collider'~\cite{Chen:2009zp,Chen:2012ge,1211.1624,Arkani-Hamed:2015bza} sensitive to new particles that might have been produced during inflation,
if their mass is around the inflationary Hubble scale $H$.
Its unknown value  can plausibly reach $H \sim 10^{13}\GeV$~\cite{CMBdata}, much beyond conventional colliders.

\smallskip

Building on the collider analogy, the `new physics' observable at the cosmological collider is non-Gaussianities in the distribution of primordial cosmological perturbations.
So far $N \sim 10^8$ modes have been observed, mostly from the Cosmic Microwave Background.
Up to $N \sim 10^{16}$ modes could be optimistically observed in future Large Scale Structure surveys.
A cubic self-coupling among inflatons, $A_\phi \phi^3$, can be probed down to $A_\phi/H \sim 1/\sqrt{N}$.
So the number $N$ of modes roughly corresponds to the number $N$ of events observed in a conventional collider in background-limited searches.
This means that the `luminosity' $N$ reached so far by the cosmological collider is already bigger than the typical 
number of highest-energy events $N = {\cal L} \sigma \sim 10^{6}$ produced at colliders such as LHC or LEP,
assuming a typical cross section $\sigma\sim 1/4\pi s$.


However, the limitation of the cosmological collider is that it collided inflatons,
at least in the simplest theory where cosmological perturbations arise from inflaton perturbations.\footnote{We do not here consider alternatives such as 
the curvaton (a scalar that after inflation dominates the energy density)~\cite{hep-ph/0109214,hep-ph/0110002, hep-ph/0110096,1908.11378}, 
modulated reheating (a scalar that controls the inflaton decay rate)~\cite{astro-ph/0303591,0709.2545,1907.07390,2112.10793}, 
couplings to chemical potentials~\cite{1805.02656,1907.10624,1908.00019,1910.12876,2004.02887,2010.04727,2203.06349}, 
tachyons~\cite{2401.11009}, non-vacuum initial states~\cite{2309.05244},
non-renormalizable derivative interactions among scalars around the EFT unitarity limit.}
The inflaton is some unknown scalar field with a known property: its potential must be nearly flat 
to drive inflation leading to the observed power-spectrum
$P_\zeta \sim \med{\zeta^2} \sim 2~10^{-9}$,
where $\zeta \sim H/M_{\rm Pl}\sqrt{\epsilon_H}$ is the curvature perturbation and $\epsilon_H \ll 1$ the slow-roll parameter.
As a result,  the inflaton is a nearly-free scalar field with small self-coupling.
Gravity contributes as an effective cubic $A_\phi  \sim \sqrt{\epsilon_H} H^2/M_{\rm Pl}$, 
suppressed by $\sqrt{\epsilon_H}$ compared to the dimensional estimate~\cite{Maldacena}. 
The overall size of 
non-Gaussianity in the probability distribution of the scalar curvature perturbation
$\zeta$ is usually parametrised as
\beq \wp (\zeta) \sim
\exp \left[ - \frac{\zeta^2}{2 \med{\zeta^2}} \left (1 + f_{\rm NL}\zeta +\cdots\right)\right]\eeq
in terms of a $f_{\rm NL} \sim \med{\zeta^3}/\med{\zeta^2}^2$ parameter.
Current data imply $\fNL \lesssim 10$~\cite{Akrami:2019izv,2404.07203}. 
Future CMB and LSS observations~\cite{Dore:2014cca} could probe $\fNL$ down to $\sim 1$~\cite{1810.13424}.
Future  21 cm tomography observations could probe  $\fNL$ down to $\sim 1$ and
maybe reach the gravitation floor of non-Gaussianity~\cite{Munoz:2015eqa,1610.06559}.
The minimal gravitational interaction corresponds to $\fNL\sim\epsilon_H \sim 10^{-2}$ and has a specific dependence on momenta~\cite{Maldacena}.
We focus on the scalar bi-spectrum, as
other correlation functions involving tensors or vectors appear less sensitive~\cite{2109.01095}. 

\smallskip

It's difficult to write simple inflation models with standard kinetic terms
that lead to large, easily observable, cosmo-collider signals.
Loop-suppressed effects are generically too small, unless strong couplings are involved.
Larger tree level  effects can arise if the scalar inflaton $\varphi$ {mixes} with an extra `isocurvaton'
scalar $\sigma$ that,
during the relevant inflation phase, has a mass around the Hubble scale $H$.
Then, on-shell production of the extra scalar during inflation imprints a
non-Gaussianity with a specific oscillatory pattern
in the wave-number $\mb{k}_{i}$ dependence of the bi-spectrum 
$\left\langle\zeta_{\boldsymbol{k}_1} \zeta_{\boldsymbol{k}_2} \zeta_{\boldsymbol{k}_3}\right\rangle$ of primordial inhomogeneities (see e.g.~\cite{Chen:2012ge,Arkani-Hamed:2015bza}).
The shape of the signal is model independent, being directly related to the kinematics of particle production,
while its amplitude depends on the couplings and, in particular, on
how fast the inflationary trajectory  `turns' in field space.


\smallskip

In section~\ref{for} we critically summarise and verify the non-trivial formalism needed to compute multi-scalar effects.
We identify the more significant contributions to $\fNL$ finding that 
(contrary to some claims in the literature) the turn rate is slow-roll suppressed
in a large class of plausible theories with Planckian physics.
As a result, in such theories, Planck-suppressed
cosmo-collider effects contribute to $\fNL$ at the level of the minimal gravitational effect, 
while affecting its wave-number dependence.

In section~\ref{scal} we compute $\fNL$ in various motivated inflationary theories with multiple scalars:
the gravi-scalar equivalent to $R^2$ gravity,
Higgs-like extra scalars.
In particular, dimension-less theories 
(where a `Planckion' scalar dynamically generates the Planck scale)
naturally lead to cosmo-collider effects.
The resulting non-Gaussianity is at the expected $\fNL \sim \epsilon_H$ level.

In section~\ref{rot} we study which inflationary models with sub-Planckian physics
allow for an enhanced turn rate,
either by having multiple constant turns during inflation allowed by special field-space geometries,
or by having a special turning feature happens around the observed $e$-fold range.

Conclusions are given in section~\ref{concl}.

\section{Formalism for two-scalar inflation}\label{for}
In this section we critically summarize the formalism that allows to compute inflation (including cosmo-collider non Gaussianities)
in theories with multiple scalar fields $\phi^a$.
We later specialise to two scalars, $a=\{1,2\}$.
We consider an Einstein-frame action of the form
\beq\label{eq:Lagen}
S =\int d^4x \sqrt{|\operatorname{det} g|} \left[
-\frac{\bar{M}_{\mathrm{Pl}}^2}{2} R+ \frac{K_{ab}(\phi)}{2}(\partial_\mu \phi^a)(\partial^\mu \phi^b)-V(\phi)\right]\eeq
where $\bp\approx 2.4~10^{18}\GeV$ is the reduced Planck mass and the kinetic metric $K_{ab}(\phi)$ describes a generic warped field space.

\subsection{Classical equations as function of time}
The classical equations of motion for spatially homogeneous fields $\phi^a_0(t)$ as function of time $t$ following from eq.\eq{Lagen}
are 
\beq
\label{eq:eomtH}
D_t \dot\phi^a_0  + 3 H \dot\phi^a_0 + K^{ab} V_b=0,\qquad\hbox{with Hubble rate}\qquad
H^2 = \frac{ K_{ab}\dot\phi^a_0 \dot\phi^b_0/2 + V}{3\bp^2}\eeq
where $V_a = \partial V/\partial\phi^a \equiv \partial_a V$, 
$K^{ab}=(K_{ab})^{-1}$, and
$D_t \dot\phi^a = \ddot\phi^a + \Gamma^a_{bc}\dot\phi^b\dot\phi^c$ is the covariant acceleration written in terms of
the
field-space Christoffel symbols
\beq \Gamma^a_{bc}=\frac12 K^{ad} \left(\frac{\partial K_{cd}}{\partial\phi^b}+ \frac{\partial K_{bd}}{\partial\phi^c}-\frac{\partial K_{bc}}{\partial\phi^{d}}\right).\eeq  
They account for centrifugal and Coriolis forces
in the case of a  flat field space written in curvilinear coordinates.
The two equations\eq{eomtH} imply $\dot H = -  \dot \phi^2_0 /2 \bp^2$
where the total speed $\dot\phi_0$ is defined as $\dot \phi^2_0 \equiv K_{ab} \dot\phi^a_0\dot\phi^b_0$.
This helps computing the first slow-roll parameter $\epsilon_H  \equiv - \dot H/H^2$.

\subsection{Classical equations as function of $e$-folds}\label{turning}
It is convenient to change variable from time to the number of $e$-folds $d{\cal N}=H\, dt$.
We denote $d/d{\cal N}$ as $'$ and $d/dt$ as a dot.
The background equations\eq{eomtH} become
\beq \label{eq:eomN}
\frac{D_{\cal N} \phi^{\prime a}_0}{3 -\epsilon_H} +  \phi_0^{\prime a} + \bp^2 \frac{K^{ab}V_b}{V} =0,\qquad
 H^2 =   \frac{V}{\bp^2(3- \epsilon_H)} \eeq
where 
$ D_{\cal N} \phi^{\prime a}= \phi''^a + \Gamma^a_{bc} \phi'^b \phi'^c$ is the covariant acceleration.
The equation for $H^2$ is obtained inserting $\dot \phi_0 = H \phi'_0$ in eq.\eq{eomtH} and solving for $H$.
The first slow-roll parameter $\epsilon_H$ becomes
\beq\qquad
\epsilon_H \equiv - \frac{H'}{H}=\frac{K_{ab} \phi^{\prime a}_0\phi^{\prime b}_0  }{2 \bp^2} = \frac{\phi'^2_0}{2\bp^2}
\eeq
where we defined the total speed $\phi'^2_0 \equiv K_{ab} \phi'^a_0 \phi'^b_0$.
It is convenient to rewrite the equations by defining a special basis that makes more transparent their geometry~\cite{astro-ph/0009131,hep-ph/0107272}.
We introduce
\begin{itemize}
\item the unit tangent vector $T^a \equiv \phi^{\prime a}_0/\phi'_0$;
\item the unit normal vector $N$ defined such that $T \cdot N=0$ and $N^2=1$
as the component of the field acceleration perpendicular to the field velocity, $ N \propto D_{\cal N} T$.
In the two-field case it is given by  $N_a \equiv\sqrt{\det K} \epsilon_{ab}T^b$ where $\epsilon_{ab}$ is the invariant anti-symmetric tensor with $\epsilon_{12}=1$.
\end{itemize}
Projecting the equations of motion\eq{eomN} over $T$ and over $N$ gives
\beq \label{eq:eqTN}
\phi''_0 + (3-\epsilon_H) \phi'_0 + \frac{V_T}{H^2} = 0 ,\qquad
 -N_a D_{\cal N} T^a = \frac{V_N}{H^2 \phi'_0}\eeq
where $V_T = T^a V_a$ and $V_N = N^a V_a$ are the potential derivatives along $T$ and $N$.
The second slow-roll parameters $\eta^a$ are defined in terms of the  covariant field acceleration,
that can be conveniently expanded in the $T,N$ basis finding
\beq \eta^a \equiv - \frac{D_{\cal N} \phi'^a_0}{\phi'_0} = \eta_T T^a + \eta_N N^a,\qquad
\eta_T = -\frac{\phi''_0}{\phi'_0} ,\qquad
\eta_N \equiv  -N_a D_{\cal N} T^a .
\eeq
The tangential component reproduces the second slow-roll parameter defined in terms
of the Hubble rate as $\eta_T = \eta_H \equiv -\epsilon_H'/2\epsilon_H$.
The normal component $\eta_N$ is known as `turn rate', 
as it describes the amount of turn per $e$-fold of the inflationary trajectory away from a geodesic in field space.
As discussed later, a large $\eta_N$ is crucial for having large non-Gaussianities.

It is convenient to additionally define the {\em curvature $\kappa$ of the inflationary field trajectory}~\cite{1010.3693}\footnote{As the formalism
might seem unusual, it is useful to point out that the  $T,N$ basis is
analogous to the standard basis for writing the velocity $\vec v$ and acceleration of a Newtonian trajectory in time. 
Eq.\eq{kappaeta} can be trivially derived from the well-known Newtonian equation $ v=\omega r$ by replacing $v\to \phi'$, $\omega\to\eta_N $, $r\to 1/\kappa$
(without confusing the osculating radius $r$ with the tensor-to-scalar ratio).}
\beq \label{eq:kappaeta} \kappa \equiv \eta_N/\phi'_0\eeq
because $\kappa$  depends only on the shape in field space of the inflationary trajectory.
The expression $\kappa = - N_a D_{\cal N} T^a/\phi'_0$ explicitly shows that $\kappa$ does not depend on the overall speed.
Thanks to this property, $\kappa$ can be simply approximated in the limit of 
vanishing slow-roll parameters, while the turn rate $\eta_N$ vanishes in the limit.
For example it's enough to approximate the inflationary trajectory as the bottom of
a nearly flat valley in field space, without having to solve the inflationary dynamics,
as replacing ${\cal N}$ with any simpler parameter along the trajectory gives the same $\kappa$.
An example will be provided in eq.\eq{kappastraight}.
This leads to an important point.
Given that $\phi' _0= \sqrt{2\epsilon_H}\bp$, the turn rate can  be large only if the curvature $\kappa$ of the inflationary trajectory is sub-Planckian
\beq \label{eq:etakappa}
\bbox{\eta_N = \phi'_0  \kappa =  \kappa\bp  \sqrt{2\epsilon_H}}.\eeq
Models that can be approximated as one-field inflation predict $r \approx 16 \epsilon_H$. 
So the turn rate tends to be even smaller in models where the scalar/tensor ratio is $r \ll 0.035$,
much below current bounds~\cite{CMBdata}.

\subsection{Quadratic action and inflationary perturbations}
Computing perturbations around the inflationary background needs a non-trivial but standard formalism, that we summarise.
We employ the Arnowitt-Deser-Misner (ADM) decomposition of metric perturbations~\cite{Arnowitt:1962hi},
\begin{align}
ds^2=-N^2 \, dt^2+h_{ij}\left(dx^i+N^i\, dt\right)\left(dx^j+N^j\, d t\right),   
\end{align}
where $i,j$ are space indices and
$N, N_i$, and  $h_{ij}$ are the lapse, shift, and induced three-dimensional metric on constant-time hyper-surfaces, respectively. 
The ADM decomposition is useful because $N$ and $N_i$ appear with no time derivatives in the action,
being gauge artefacts corresponding to $x^\mu\to x^\mu + \xi^\mu(x)$ reparametrizations.
Focusing on scalar perturbations, we expand around the inflationary FLRW background as
\begin{align}
N=1+\alpha, \qquad N_i=\partial_i \beta, \qquad h_{ij}=a^2(t) \delta_{ij},    
\end{align}
where we choose the spatially flat gauge and omitted the vector and tensor perturbations.
Next, the scalars $\phi^a$ are expanded in perturbations $\varphi^a$ as  
\begin{align}
\phi^a=\phi_0^a+\varphi^a - \frac{1}{2} \Gamma_{bc}^a \varphi^b \varphi^c+\cdots.     
\end{align}
where  $\phi_0(t)$ is the classical inflationary background discussed in the previous subsections.
The Christoffel terms define $\varphi$ as the deviation from geodesic motion, 
such that the perturbations are covariant under field-space reparametrizations~\cite{Gong:2011uw,1208.6011,2309.10841}.
Inserting the metric and the field expansions into eq.~\eqref{eq:Lagen} 
gives the action for the perturbations $\varphi^a,\alpha,\beta$.

\smallskip

Expanding the action up to quadratic order 
confirms that the metric perturbations $\alpha$ and $\partial_i^2 \beta$ appear as auxiliary fields without time derivatives,
as expected  due to reparametrization invariance.
This allows to integrate out $\alpha$ and $\beta$ as~\cite{Maldacena,Gong:2011uw}
\beq \label{eq:alphabeta}
\alpha =\epsilon_H  \frac{H}{\dot\phi_0} T_a \varphi^a,\qquad
\frac{\partial_i^2 \beta}{a^2} = - \frac{\alpha (6\bp^2 H^2 - \dot\phi_0^2) + \dot\phi_0 T_a \, D_t\varphi^a + V_a \varphi^a}{2H \bp^2}\eeq
where $\dot{\phi}_0^2\equiv K_{ab} \dot{\phi}_0^a \dot{\phi}_0^b$.
The resulting quadratic action for scalar field fluctuations $\varphi^a$ is
\beq \label{eq:S2flat}
S^{(2)}=\int d^4x~ a^3 \left[ \frac{K_{ab}}{2} (D_t \varphi^a)(D_t \varphi^b)  - \frac{K_{ab}}{2a^2} (\partial_i\varphi^a)(\partial_i\varphi^b) - \frac{M^2_{ab}}{2}\varphi^a \varphi^b\right].\eeq
It contains no $\varphi\, D_t \varphi$ terms.
The scalar squared  mass matrix is~\cite{0806.0336,0806.0336,Gong:2011uw,1208.6011,2309.10841,1606.06971}
\beq \label{eq:MMab}
M^2_{ab}=V_{ab} + \frac{2 \epsilon_H H}{\dot\phi_0}(V_a T_b + V_b T_a) +
2\epsilon_H (3-\epsilon_H) H^2 T_a T_b  +  \dot\phi_0^2 {\cal R}_{aTbT}\eeq
where $ {\cal R}_{aTbT} \equiv T^c T^d {\cal R}_{acbd} $.
The latter two contributions arise from the gravitational part of the action, after eliminating $\alpha,\beta$.
We define the derivatives of the potential as
\beq \label{eq:dV} V_a = D_a V = \partial_a V,\qquad
V_{ab}=D_b D_a V,\qquad V_{abc}=D_c D_b D_a V.\eeq 
In view of the covariant derivative $D_a$,
the third derivative $V_{abc}$ is not symmetric.

\smallskip

Specializing to two scalars only, 
the Riemann tensor has one component and 
can be written in terms of the field-space curvature ${\cal R}$  as
${\cal R}_{abcd} = {\cal R} (K_{ac}K_{bd}-K_{ad}K_{bc})/2$.
Furthermore, the quadratic action gets almost diagonalised by decomposing the scalar fluctuations $\varphi^a$ along the
tangent $T^a$ and normal direction $N^a$ as
\begin{align}\label{eq:TNbasis}
\varphi^a=\varphi\, T^a+\sigma \, N^a,    
\end{align}
where $\varphi$ and $\sigma$ are known as the {\em adiabatic} (or inflaton) mode and  the {\em isocurvature} mode, respectively. 
Then kinetic terms induce extra mass terms, as $T^a$ and $N^a$ are time-dependent.

The action of eq.\eq{S2flat}, expressed in terms of $\varphi$ and $\sigma$, becomes~\cite{0806.0336,0806.0336,1010.3693,1606.06971,2405.11628}
\begin{eqnarray}
\label{eq:S2varphigauge}
S^{(2)} &=& 
 \int d^4x~a^3 \bigg[
\left( \frac{\dot\varphi^2}{2} - \frac{(\partial_i \varphi)^2}{2a^2}  - \frac{m_\varphi^2}{2}\varphi^2\right)
+\left( \frac{\dot\sigma^2}{2} - \frac{(\partial_i \sigma)^2}{2a^2}  - \frac{m_\sigma^2}{2}\sigma^2\right) +
\Lag_{\varphi\sigma}\bigg],\\
\Lag_{\varphi\sigma}  &=& \eta_NH (\sigma\dot\varphi - \dot\sigma\varphi)  - (V_{NT} + 2\epsilon_H H^2 \eta_N) \sigma\varphi
=2\eta_N  (H \sigma\dot\varphi +  \eta_H H^2 \sigma\varphi).
\label{eq:Lvarphisigma}
\end{eqnarray}
In the last expression we integrated by parts the Coriolis term;
this generates mass mixing terms that cancel $V_{NT}$, leaving
terms suppressed by slow-roll parameters. The mass terms are
\begin{eqnarray}
\label{eq:m_sigma}   
m_{\sigma}^2 &=&  M^2_{NN} -H^2\eta_N^2=
 V_{NN} + H^2 \left( \frac{\phi'^2_0}{2}{\cal R}-\eta_N^2\right),\\
 m_\varphi^2 &=& M^2_{TT} - H^2\eta_N^2 = V_{TT}+2\epsilon_H H^2 (3-\epsilon_H)
+4 \epsilon_H H V_T/\dot\phi_0 - H^2\eta_N^2.
\end{eqnarray}
The isocurvaton mass $m_\sigma^2$ can receive a significant contribution from
$V_{NN}\equiv N^a N^b  V_{ab}$,
while derivatives of the potential involving the inflaton $T$ direction are suppressed by 
slow-roll parameters:
\begin{eqnsystem}{sys:VT}
\label{eq:VTT}
V_{TT} &\equiv & T^a T^b  V_{ab}= H^2 [\epsilon_H (6-2\epsilon_H-5\eta_H) +  \eta_H(3-\eta_H + \eta_{H2}) + \eta_N^2],\\
\label{eq:VNT}
V_{NT} &\equiv & N^a T^b  V_{ab} = H^2 \eta_N (3-3\epsilon_H - 2 \eta_H+\eta_{N2}).
\end{eqnsystem}
The slow-roll parameters are defined as
\beq \label{eq:SRs}
\epsilon_H = - \frac{d\ln H}{d{\cal N}},\qquad
\eta_H = -\frac12 \frac{d\ln \epsilon_H}{d{\cal N}},\qquad
\eta_{H2}=\frac{d \ln \eta_H}{d{\cal N}},\qquad
\eta_{N2}=\frac{d \ln \eta_N}{d{\cal N}}.
\eeq
As a result the inflaton mass $m_\varphi$ is small but non-vanishing~\cite{1010.3693,1606.06971,2405.11628}.
As discussed later, $m_\varphi$ and the mass mixing term in $\Lag_{\varphi\sigma}$ are sub-leading gauge artefacts.



\subsubsection*{Power spectrum from the slow-roll $\delta{\cal N}$ formalism}
Cosmological data probe the curvature perturbation $\zeta$, related at leading order to the inflaton fluctuation $\varphi$ by
\beq \label{eq:zetavarphi} \zeta = - H \,\delta t = - H \frac{\varphi}{\dot\phi_0}.\eeq
Indeed $\zeta$ is defined as the gravitational scalar perturbation $h_{ij} = a^2 e^{2\zeta} \delta_{ij}$
in the coordinate choice where the inflaton is used as clock, such that $\varphi=0$.
So $\zeta$  describes perturbations in the curvature of fixed-time 3-dimensional spaces.
The two coordinate choices are connected by redefining the time-coordinate $t\to t + \delta t$, obtaining eq.\eq{zetavarphi}.

The power spectrum in multi-field inflation $P_\zeta(k)$
can be computed in the Gaussian limit by solving
the Mukhanov-Sasaki equations for the fluctuations following from the quadratic action of eq.\eq{S2varphigauge} (see e.g.~\cite{1502.03125}).
For our purposes it is enough to use the $\delta{\cal N}$ formalism, that relates
the curvature fluctuation $\zeta$ to the number of 
$e$-folds ${\cal N}$ up to inflation end (not to be confused with the normal $N$)
and expanding it in terms of scalar fluctuations~\cite{astro-ph/9507001,astro-ph/9604103,0810.5387}
\beq \zeta = \delta {\cal N} = {\cal N}_a   \varphi^a + \cdots.\eeq
The $\delta{\cal N}$ formalism holds assuming that field velocities can be expressed in terms of field values rather than being independent variables.
This is true if the equations of motion are of first order. Such approximation holds in the slow-roll limit,
assuming that terms with second derivatives and squared first derivatives can be neglected 
(so the curvature cannot be too large).
The classical equation of motion of eq.\eq{eomN} and the Hubble slow-roll parameter simplify to
\beq
\frac{d\phi^a}{d{\cal N}}  \stackrel{\rm SR}{\simeq}   - \frac{\bp^2}{V} K^{ab}\frac{\partial V}{\partial \phi^b},\qquad
\epsilon_H \stackrel{\rm SR}{\simeq} \epsilon_V\equiv  \frac{\bp^2}{2}  \frac{K^{ab} V_a V_b}{V^2} .
\eeq
The scalar power spectrum is
\beq 
P_\zeta = \left(\frac{H}{2\pi}\right)^2 (\nabla {\cal N})^2,\qquad
(\nabla {\cal N})^2\equiv  K^{ab} \frac{\partial {\cal N}}{\partial \phi^a}\frac{\partial {\cal N}}{\partial \phi^b}.\eeq
The tensor power spectrum is  $P_t = ({2}/{\bp^2}) \left(\sfrac{H}{2\pi}\right)^2 $ so
\beq r \equiv \frac{4P_t}{P_\zeta} =\frac{8}{\bp^2(\nabla {\cal N})^2}.
\eeq
The scalar spectral index is computed as~\cite{astro-ph/9507001,astro-ph/9604103,0810.5387}
\beq n_s - 1 \equiv \frac{d\ln P_\zeta}{d\ln k}= 2\frac{\dot H}{H^2} + 2   \frac{{\cal N}^a {\cal N}^b}{(\nabla {\cal N})^2} \frac{M^2_{ab}}{3H^2}\eeq
in terms of the squared mass term of eq.\eq{MMab} that couples
the Mukhanov-Sasaki  equations for the time evolution of the field fluctuations $\varphi^a$.
Expanding in the slow-roll potential parameters $\epsilon_V$ and $\eta_{ab} = \bp^2 V_{ab}/V$ gives
\beq n_s - 1 =
-{2\epsilon_V}+ {2\eta_{ab} \frac{  {\cal N}^a {\cal N}^b }{(\nabla {\cal N})^2 }  }
{-\frac{2 }{(\nabla {\cal N})^2\bp^2 }}
+\bp^4 \frac{2 {\cal N}_a {\cal N}^b {\cal R}^a{}_{cbd} V^b V^d}{3(\nabla {\cal N})^2 V^2}
\eeq
where ${\cal R}_{acbd} $ is the Riemann tensor in field space.
The $\Lambda$CDM best-fit values around the CMB scale $k_{\rm CMB}\approx 0.056/{\rm Mpc}$
corresponding to ${\cal N}\approx 50-60$ $e$-folds before the end of inflation are~\cite{CMBdata}
\beq \label{eq:infpredex}
n_s = 0.9649 \pm 0.0042 ,\qquad P_\zeta(k_{\rm CMB}) \approx (2.1 \pm 0.06) \times 10^{-9},\qquad
r\lesssim 0.035.\eeq

\subsection{Cubic action in the flat $\varphi$-gauge}
The $\delta{\cal N}$ formalism provides an approximation for non-Gaussianities,
$\fNL \simeq 5 {\cal N}_{ab} {\cal N}^a {\cal N}^b/6(\nabla {\cal N})^2$,
in the local limit where  scalars are much lighter than the Hubble scale, see e.g.~\cite{2212.14035}.
We are instead interested in the cosmo-collider regime $m_\sigma \sim H$.
The non-Gaussianity can be computed perturbatively expanding in the
 quadratic mixing interaction $\Lag_{\sigma\varphi}\simeq 
2\eta_N H \sigma\dot\varphi$
(while the additional mass mixing term can be neglected)
and in the couplings in the higher-order action.
Expanding the action up to cubic order gives, in the flat gauge~\cite{Gong:2011uw,1208.6011,1708.07130,2309.10841} 
\begin{eqnsystem}{sys:L3flat}
S^{(3)} &=& \int d^4 x \, a^3 \left[\Lag_{\rm gravity}^{(3)} + \Lag_{\rm mix}^{(3)}+\Lag_{\rm scalars}^{(3)} \right] \\
\riga{up to total derivatives and terms proportional to the classical equations of motion.
The first term is the standard purely gravitational cubic in the metric perturbations $\alpha,\beta$:}\\[-1ex]
\Lag_{\rm gravity}^{(3)} &=& \alpha^3 (3\bp^2 H^2 - \dot\phi_0^2/2) + 2\bp^2 H \, \alpha^2 \frac{\partial_i^2 \beta}{a^2}
+ \alpha \frac{\bp^2}{2}  \frac{ (\partial_i^2 \beta)^2 - (\partial_i \partial_j \beta)^2}{a^4}.\\
\riga{The second term contains cubics that involve both $\alpha,\beta$ and scalars $\varphi^a$:}
\Lag_{\rm mix}^{(3)} &=& -\frac{\alpha}{2} \bigg[(D_t \varphi^a)(D_t \varphi_a) + \frac{(\partial_i \varphi_a)(\partial_i \varphi^a)}{a^2}
+(V_{ab} + \dot\phi_0^2 {\cal R}_{aTTb})\varphi^a\varphi^b
\bigg]+ \nonumber
\\
&&+\alpha^2 \dot\phi_0^{a} D_{t} \varphi_{a}  +\frac{\partial_i \beta}{a^2}\left[ \alpha \dot\phi_0 \partial_i \varphi-
(\partial_i \varphi_a)(D_t\varphi^a)\right].\\
\riga{The third term contains cubic couplings among scalars:}\\
\Lag_{\rm scalars}^{(3)} &=& \frac{1}{6}(- V_{abc}+{\cal R}_{adeb;c} \dot\phi_0^d \dot\phi_0^e)
\label{eq:L3scalflat}
 \varphi^a \varphi^b \varphi^c + \frac{2}{3}{\cal R}_{abcd}(D_t \varphi^a) \varphi^b \varphi^c \dot{\phi}_0^d.
\end{eqnsystem}
The covariant field expansion recognises that field-dependent kinetic terms only
contribute proportionally to the Riemann tensor.\footnote{For example~\cite{Chen:2009zp}
considered a flat kinetic metric in polar coordinates $\phi_{1,2}$, such that the term
$\phi_1^2 (\partial\phi_2)^2$ gives a cubic $\varphi_1 (\partial \varphi_2)^2$,
which gets cancelled in the covariant field expansion, as it recognises that polar coordinates describe flat space.
This issue is numerically non important in~\cite{Chen:2009zp}, where the different cubic $V_{NNN}$ gives the dominant contribution to $\fNL$. }

\smallskip

The first two terms, $\Lag_{\rm gravity}^{(3)} $ and $\Lag_{\rm mix}^{(3)} $,  can be neglected, as they lead to cosmo-collider effects suppressed
by extra slow-roll parameters compared to the scalar cubics in $\Lag_{\rm scalars}^{(3)}$.
Indeed, specialising eq.\eq{alphabeta} to two fields, $\alpha$ and $\beta$ are given by  
\beq \alpha = \frac{\epsilon_H}{\phi'_0} \varphi,\qquad\hbox{and}\qquad 
 - \frac{\partial_i^2\beta}{a^2} = \frac{(6\bp^2- \phi'^2_0)H^2 \alpha + H^2  \phi'_0 \varphi'  + V_T \varphi + 2 V_N \sigma}{2 H \bp^2}.
\eeq
So $\alpha$ only depends on the inflaton $\varphi$, 
and both $\alpha$ and $\beta$ are suppressed by slow-roll parameters, since
\beq\label{eq:VTVN}
\phi'_0 = \sqrt{2\epsilon_H}\bp,\qquad
V_T =  -H^2 \phi'_0 (3-\epsilon_H - \eta_H),\qquad
V_N = \eta_NH^2 \phi'_0. \eeq
The cubic scalar term leads to the least suppressed contributions to $\fNL$.
As long as the Riemann tensor in eq.\eq{L3scalflat} is Planck-suppressed,
the main contribution to scalar cubic interactions comes simply from the  third derivative $V_{abc}$
of the potential defined in eq.\eq{dV} and
projected along the inflaton and isocurvaton directions
\begin{align}\label{eq:Vabc}
-\frac{1}{6}V_{abc} \varphi^a \varphi^b \varphi^c=-\frac{V_{NNN}}{6}\sigma^3-\frac{V_{NNT}}{2}\sigma^2\varphi-\frac{V_{NTT}}{2}\sigma \varphi^2+\cdots,
\end{align}
where 
\beq 
V_{NNN} = N^a N^b N^c V_{abc},\qquad
V_{NNT} = N^{(a} N^b T^{c)} V_{abc}, \qquad
V_{NTT} = N^{(a} T^b T^{c)} V_{abc}\eeq
and parentheses denote symmetrisation, e.g.\ $N^{(a} N^b T^{c)} = (N^a N^b T^c + N^b N^c T^a + N^c N^a T^b)/3$.


\begin{figure}[t]
\begin{center}
$$\includegraphics[width=\textwidth]{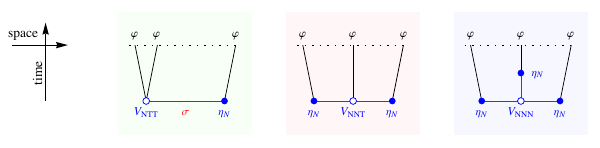}$$
\caption{\label{fig:Feyn}\em Feynman diagrams for computing the non-Gaussian bispectrum of the adiabatic inflaton mode $\varphi$ mediated at tree level by an isocurvaton scalar $\sigma$ (in red),
coupled via a mixing term proportional to the turn rate $\eta_N$ and by various cubic couplings.}
\end{center}
\end{figure}

\begin{figure}[t]
\begin{center}
$$
\includegraphics[width=0.48\textwidth]{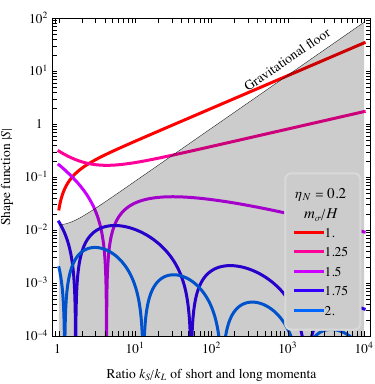}
\qquad
\includegraphics[width=0.45\textwidth]{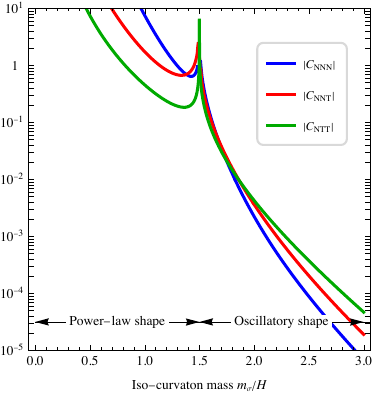}
$$
\caption{\label{fig:fC}\em 
{\bfseries Left panel:} gravitational floor (in gray) and sample
cosmo-collider shape functions of eq.\eq{S_func} in the squeezed limit $k_L\ll k_S$,
assuming the $\fNL$ discussed below eq.\eq{fNLVNTT}.
{\bfseries Right panel:}  Feynman diagram functions of eq.~(\ref{sys:Cs}).
The peak in $\fNL$ at $m_\sigma /H = 3/2$ does not affect the shape function, as shown in the left panel (see also~\cite{1811.00024,2012.13667,2112.05710}).}
\end{center}
\end{figure}






\subsection{Non-Gaussianity in the flat $\varphi$-gauge}
The measured bispectrum of the curvature perturbation $\left\langle\zeta^3\right\rangle$ is frequently parameterized by the dimensionless {\em shape function} $S$ as
\begin{align}
\left\langle\zeta_{\boldsymbol{k}_1} \zeta_{\boldsymbol{k}_2} \zeta_{\boldsymbol{k}_3}\right\rangle=\frac{(2 \pi)^7 P_\zeta^2  }{\left(k_1 k_2 k_3\right)^2}\ 
\delta\left(\mb{k}_1+\mb{k}_2+\mb{k}_3\right)
S\left( \frac{k_1}{ k_3}, \frac{k_2}{k_3}\right) \, .   
\end{align}
We are interested in cosmological collider signals that appear 
in the squeezed limit where one of the momentum is  smaller than the others, 
$k_3 \ll k_1\simeq k_2$, so that the momenta are better
denoted as short and long, $k_3 \equiv k_L \ll k_{1,2}\equiv k_S$. 
In the squeezed limit, the shape function is approximately given by (see Appendix~\ref{calculation of bispectrum} for details)
\beq
S \simeq \frac{9}{10}
\left[ f_{\rm NL}(\nu) \left(\frac{k_L}{k_S}\right)^{1 / 2-\nu}   \!\!\!\! +
 f_{\rm NL}(-\nu)   \left(\frac{k_L}{k_S}\right)^{1 / 2+\nu}\right]\quad\hbox{where}\quad
\nu = \sqrt{\frac94- \frac{m_\sigma^2}{H^2}}.
 \label{eq:S_func}  
\eeq
When $m_\sigma \sim H$ the shape function acquires a non-analytic form characteristic of cosmological collider signals:
\begin{itemize}
\item For $m_\sigma/H<3/2$ the parameter $\nu$ is real; the second term in eq.~\eqref{eq:S_func}
can give a growing power, while the  first term is subdominant.
\item For $m_\sigma/H>3/2$ the parameter  $\nu$ is purely imaginary,
giving a more characteristic oscillating shape $(k_L/k_S)^{1/2\pm \nu}$ but no enhancement at $k_S\ll k_L$.
Compared to the previous case, larger values of $\fNL$ are thereby needed to have an
oscillating signal above the gravitational floor, that contributes as~\cite{Maldacena,1106.1462}
\beq S|_{\rm grav} \simeq  \frac{1-n_s + {\cal O}({k_L}/{k_S})^2 }{4} \frac{k_S}{k_L} \qquad\hbox{in the squeezed limit $k_L\ll k_S$.}\eeq
\end{itemize}
These features are illustrated in the left panel of fig.\fig{fC}.

\smallskip

We compute the contribution to $\med{\varphi^3}$ from the dominant
non-derivative couplings in eq.\eq{Vabc},
and naively convert it into $\med{\zeta^3}$ by using the leading order relation of eq.\eq{zetavarphi}.
This approximation captures the leading effect.
Then, the three Feynman diagrams in fig.\fig{Feyn} contribute at tree level to $\fNL$ with different powers of the turn rate $\eta_N$ and same shape $S$:
\beq \label{eq:fNLcc}
\fNL=\frac{10}{9\sqrt{P_\zeta}}\left[\eta_N  C_{NTT} \frac{V_{NTT} }{H}+
\eta_N^2  C_{NNT} \frac{V_{NNT} }{H}+\eta_N^3  C_{NNN} \frac{V_{NNN} }{H} \right].\eeq
The dimension-less functions $C$ of the iso-curvature mass in eq.~\eqref{eq:fNLcc} are computed in Appendix~\ref{calculation of bispectrum}, finding
\begin{eqnsystem}{sys:Cs}
\label{eq:CNTT}
C_{NTT}(\nu) &=& \frac{2^{2 \nu-4}(2 \nu-5)}{\sqrt{\pi}(3+2 \nu)} \Gamma(\nu) \Gamma\left(\frac{1}{2}-\nu\right) \operatorname{tan}\left[\frac{\pi}{4}(1-2 \nu)\right],\\
C_{NNT}(\nu) &=&  \frac{-2^{\nu-4}\sqrt{\pi}\Gamma(\nu) }{\cos\left(\frac{\pi \nu}{2}\right)+\sin \left(\frac{\pi \nu}{2}\right)} \int_0^{\infty} d z \, z^{-\frac{5}{2}-\nu} \operatorname{Im}\left[(1+i z) e^{-i z} \mathcal{I}_{+}(z)\right],\\
C_{NNN}(\nu)&=&  \frac{-2^{\nu-5}\pi^{3/2}\Gamma(\nu) }{\cos\left(\frac{\pi \nu}{2}\right)+\sin \left(\frac{\pi \nu}{2}\right)} \int_0^{\infty} d z \, z^{-\frac{5}{2}-\nu} \operatorname{Im}\left[\mathcal{I}_{+}^2(z)\right], 
\end{eqnsystem}
where the function $\mathcal{I}_{+}(z)$ is given in eq.~\eqref{I_func}.
The $C$ functions are numerically plotted in fig.\fig{fC}.

\medskip

Cosmo-collider effects arise if $m_\sigma \sim H$, a regime 
known as quasi-single field inflation, as the isocurvaton $\sigma$ is heavy enough that it 
stays around the bottom of its potential, and light enough to produce cosmo-collider signals.
For $m_\sigma \gg H$ the $C$ functions get Boltzmann-suppressed as $e^{-\pi m_\sigma/H}$,
while for small masses $m_\sigma \ll H$ the shape function becomes analytic, and
the IR divergence  is cut by 
the finite duration of inflation, becoming an enhancement by ${\cal N}\sim 60$~\cite{Chen:2009zp}. 

Eq.\eq{m_sigma} for $m_\sigma$ implies (ignoring tunings) that $m_\sigma \sim H$ can arise at
$|\eta_N |\lesssim 1$ and for field-space  curvature  $|{\cal R}| \lesssim 1/\bp^2\epsilon_H$.
The only large contribution to $\fNL$ can arise from $V_{NNN}$ at cubic order in the turn rate $\eta_N$.
Indeed the other potential derivatives $V_{NTT}$ and $V_{NNT}$ that involve the inflaton direction $T$ 
can be expressed in terms of $V_{NN}$ and of slow-roll parameters in eq.\eq{SRs} as\footnote{Eq.\eq{VNTTexpanded}
is found by differentiating eq.\eq{VNT} with respect to ${\cal N}$;
$V_{NN}$ arises from $D_{{\cal N}}T = - \eta_N N$.
Eq.\eq{VTTTexpanded} arises by similarly differentiating eq.\eq{VTT}.
Eq.\eq{VNNTexpanded} arises by differentiating $V_{NN}$.
}
\begin{eqnsystem}{sys:V'''}
 \label{eq:VNTTexpanded}
 V_{NTT} &=&  \frac{\eta_N}{{\phi}'_0} \bigg[V_{N N} + H^2
(-12\epsilon_H - 3\eta_H+3\eta_{N2}+ {\cal O}(\epsilon^2) )
\bigg]
-  \eta_N\frac{\mathcal{R}}{6} H^2 \phi_0^{\prime},\\
\label{eq:VTTTexpanded}
V_{TTT} &=& \frac{H^2}{\phi'_0} \bigg[6\eta_N^2 -12\epsilon_H^2+ 3\eta_H \eta_{H2}-18\eta_H \epsilon_H + {\cal O}(\epsilon^3)\bigg],\\
\label{eq:VNNTexpanded}
V_{N N T} &=& \frac{{V}'_{NN}}{{\phi}'_0}-\frac{6 H^2\eta_{N}^2}{{\phi}'_0} (1+{\cal O}(\epsilon))
- \mathcal{R}  H^2 \phi_0^{\prime} \left(1-\frac{\epsilon_H+\eta_H}{3}\right).
\end{eqnsystem}
Unless ${\cal R}$ is much larger than Planckian,
the only possibly large contribution arises from the $V_{NN}$ contribution to $V_{NTT}$.
However the cosmo-collider regime $m_\sigma \sim H$ restricts $V_{NN}\sim H^2$.
As a result the $V_{NN}$ term in $V_{NTT}$ is dominant, but still relatively small.
In conclusion, while $\fNL$ receives many contributions, only the two following contributions seem interesting:
\begin{itemize}
\item A large $\fNL$ can arise from triple-isocurvaton exchange
in special models where $\eta_N$ and $V_{NNN}$ are large~\cite{Chen:2009zp}.

\item Single-isocurvaton exchange with the $V_{NTT}$ interaction contributes as
\beq \label{eq:fNLVNTT}
\fNL \simeq 2\pi \frac{10}{9} \eta_N^2 C_{NTT}  \frac{V_{NN}}{H^2}\qquad\hbox{($\varphi$ gauge)}.   \eeq
having used $\sqrt{P_\zeta} =(H/\bp)/\sqrt{8\pi^2 \epsilon_H}$ and $\phi'_0 = \sqrt{2\epsilon_H}\bp$.
Eq.\eq{etakappa} shows that
the turn rate is $\eta_N^2\sim\epsilon_H$
in models with Planckian-size curvature of field space and 
potentials that allow at most one turn during inflation.
The resulting $\fNL\sim\epsilon_H$
is comparable to the gravitational contribution, but has a distinctive cosmo-collider shape
that might help its detectability~\cite{1610.06559}.
The left panel of fig.\fig{fC} illustrates the shape function corresponding to eq.\eq{fNLVNTT} for $\eta_N=0.2$ and $V_{NN}=H^2$.
\end{itemize}
This will be confirmed by explicit model computations in section~\ref{scal}, 
and in section~\ref{sec:zeta} by a computation done in an alternative gauge.

\subsection{Action and non-Gaussianity in the comoving $\zeta$-gauge}\label{sec:zeta}
The flat $\varphi$-gauge produces gauge artefacts on super-horizon scales: 
the scalar $\varphi$ has a small mass term and non-derivative interactions; so it keeps evolving after horizon exit.
A different coordinate choice is more convenient after horizon exit:
the comoving gauge where $\varphi$ is set to 0 and gets substituted by the curvature perturbation $\zeta$, see eq.\eq{zetavarphi}.
The comoving $\zeta$-gauge is convenient because $\zeta$ does not evolve outside the horizon,
showing that inflationary predictions do not depend on the unknown physics after inflation ends.
Technically this happens because $\zeta$ is massless and only has derivative interactions.
Such property can be understood noticing that $\zeta$, 
being as in eq.\eq{zetavarphi} the clock telling the time delay between universes separated by inflation,
is the Goldstone boson of broken time translation invariance, see e.g.~\cite{0709.0293,1606.06971}
and~\cite{1303.1523} for a review.
However, the $\zeta$-gauge produces artefacts on sub-horizon scales, such as misleading estimates of $\fNL$.
Technically, these can be avoided by rewriting the $\zeta$-gauge action performing laborious integrations by parts~\cite{Pinol}.
We ignore boundary terms, although some total time derivatives need extra considerations~\cite{Pinol,2403.14558,2403.16022,1103.4126,1103.1102,1104.0292}. 
In the comoving gauge the gravitational terms are given by~\cite{1303.1523,Pinol} 
\beq \alpha =\frac{\dot\zeta}{H} 
,\qquad
 \beta =  -\frac{\zeta}{H} 
+\chi,\qquad
\frac{\partial_i^2 \chi}{a^2}=
\epsilon_H\dot\zeta -  \eta_N \frac{ \dot\phi_0}{\bp^2}\sigma .
\eeq
This again implies that the gravitational part of the action contributes to slow-roll suppressed $\fNL$.
At quadratic level, $\zeta$ couples to the isocurvature mode $\sigma$ only by
the `Coriolis force' term proportional to the turn rate 
\begin{align}
\label{eq:S2zetagauge}
S^{(2)}= \int d^4 x~a^3  \left[\frac{\dot{\phi}_0^2}{2H^2}\left(\dot{\zeta}^2-\frac{(\partial_i \zeta)^2}{a^2}\right)+
\frac{1}{2}\left(\dot{\sigma}^2-\frac{(\partial_i \sigma)^2}{a^2}-m_\sigma^2 \sigma^2\right)-2 \dot{\phi}_0 \eta_N \dot{\zeta} \sigma\right]
.
\end{align}
The isocurvaton squared mass $m^2_{\sigma}$ is given by eq.\eq{m_sigma}, while $\zeta$ is massless.
The above action holds without assuming the slow-roll approximation.
Ignoring total derivatives and terms proportional to the classical equations of motion for fluctuations $\zeta,\sigma$, 
the cubic action is~\cite{Pinol}
\begin{eqnsystem}{sys:L3comoving}
S^{(3)} &=&  \int d^4 x~a^3  \bigg[ \Lag^{(3)}_{\rm gravity}+ \Lag^{(3)}_{\rm mix} +
 \Lag_{\sigma\zeta\zeta} + \Lag_{\sigma\sigma\zeta} + \Lag_{\sigma\sigma\sigma} 
\bigg] , \\
\label{eq:Lagszz}
\Lag_{\sigma \zeta\zeta}  & = &  
\frac{\dot\phi_0\eta_N}{H} \sigma \left[ \dot\zeta^2  -\frac{(\partial \zeta)^2}{a^2}\right]  -
(2 \epsilon_H - \eta_N +2\eta_H - 2\eta_{N2}) \dot\phi_0 \eta_N \zeta \dot\zeta \sigma , \\
 \Lag_{\sigma\sigma\zeta} &=& \frac{m_\sigma^2}{2}(\epsilon_H + \eta_\sigma) \zeta\sigma^2
 - H (\eta_N^2 -\epsilon_H \bp^2 {\cal R}) \dot\zeta \sigma^2
+\frac{\epsilon_H}{2}\zeta \left[\dot\sigma^2 + \frac{(\partial_i \sigma)^2}{a^2}\right], \\
\Lag_{\sigma\sigma\sigma}  &=& - \frac16 (V_{NNN} +2\dot\phi_0 H \eta_N {\cal R} + \epsilon_H H^2 \bp^2 {\cal R}_T) \sigma^3
\label{eq:L3scalcomoving}
\end{eqnsystem}
where $\eta_\sigma\equiv d\ln m_\sigma^2/d{\cal N}$.
We omitted $\Lag^{(3)}_{\rm gravity}$ and $\Lag^{(3)}_{\rm mix} $ 
as they again give the usual single-field  `gravitational floor' contribution, $\fNL\sim \epsilon_H$, plus
sub-leading effects~\cite{Maldacena,Pinol}.
We recover the two main contributions to $\fNL$ discussed at the end of the previous section:
\begin{itemize}
\item Triple-isocurvaton exchange proportional to $V_{NNN}$.

\item Single-isocurvaton exchange. 
The $\varphi$-gauge contains  a
$V_{NTT} \sigma \varphi^2 $ non-derivative interaction, resulting in the $C_{NTT}$ loop function.
The $\zeta$-gauge action contains, instead of $V_{NTT}$, a derivative interaction
(dominant first term in eq.\eq{Lagszz}) that leads to a different $\tilde{C}_{NTT}$ loop function.
The resulting
\beq \label{eq:fNLzeta}
\fNL =\frac{10}{9} \eta_N^2 \tilde{C}_{NTT},\qquad
\tilde{C}_{NTT} =2\pi \left( \frac{m_\sigma}{H}\right)^2
C_{NTT}\qquad \hbox{($\zeta$ gauge)}\eeq
agrees with the $\varphi$-gauge result in eq.\eq{fNLVNTT}, up to sub-leading terms,
after taking into account that $m_\sigma^2 \simeq V_{NN}$, eq.\eq{m_sigma}.
\end{itemize}
Having summarized the formalism and checked its validity, we next perform model computations.

\section{Inflationary theories with multiple scalars}\label{scal}
The exploration of inflation theories with two scalars runs the risk of becoming an extensive list of non-predictive models.
We start focusing on motivated scalars with motivated potentials.
Observational data exclude most large-field inflation models, favouring theories with an especially flat inflaton potential 
such that the tensor/scalar ratio $r$ satisfies current upper bounds.
Especially flat potentials generically arise within dimension-less theories, that thereby are power-counting renormalizable.
Two special examples are:
 \begin{itemize}
\item[H)]  {\em Higgs-like inflation}, driven by a
scalar $s$ with quartic potential $\lambda_s s^4/4$ and non-minimal coupling to gravity 
$\xi_s s^2 R$, where $\lambda_s$ and $\xi_s$ are dimension-less couplings. 
The quasi-flat potential arises when writing the action in the Einstein frame.

\item[S)] {\em Starobinsky inflation}, with action $S = \int d^4 x \sqrt{|\det g|} [ - \bp^2 R/2+R^2/6 f_0^2]$, where $f_0$ is a dimension-less coupling.
When rewritten in the Einstein frame, the extra degree of freedom present in the 4-derivative gravitational term $R^2$
is equivalent to an extra scalar (here named as `gravi-scalar' and denoted as $z$) with a quasi-flat potential.
\end{itemize}
In both cases the flatness is lost if, following the Effective Field Theory logic,
 extra non-renormalizable terms (such as $s^5$ or $R^3$) are added to the action with sizeable coefficients.
 For example, Starobinsky inflation does not arise naturally from  string theory, 
 as the string scale is dimensionful~\cite{2312.13210}.
 
The cosmo-collider probes extra scalars with inflationary mass comparable to the Hubble scale.
Such a coincidence naturally arises in dimension-less theories with comparable couplings.
Furthermore, dimension-less theories tend to maximise cosmo-collider effects (up to tunings and/or special structures).
Indeed, large cosmo-collider effects need large cubics $V'''$, that are however limited
by the requirement that the adiabatic and the isocurvature mode are light enough, $V'' \lesssim H^2$:
in dimension-less theories both $V''$ and $V'''$ arise from quartic dimension-less couplings, 
roughly saturating the naturalness bound $V''' \sim V''/\phi$.

\medskip

We will consider theories with generic scalars $\phi$ and Jordan-frame action\footnote{The most general renormalizable
action also contains a $R_{\mu\nu}^2$ term, that gives rise to a problematic spin-2 ghost,
and to negligible loop-level cosmo-collider effects in $\med{\zeta^3}$.
Cosmo-collider ghost effects at tree-level can arise in $\med{\zeta^4}$~\cite{2210.16320} and in
correlation functions involving spin-2 modes $\med{h_{\mu\nu} \zeta^2}$~\cite{1607.03735}.
We focus on the gravi-scalar contained in $R^2$:
having spin 0 it leads to tree-level cosmo-collider signals in $\med{\zeta^3}$.
}
\beq \label{eq:S1}
S = \int d^4 x \sqrt{|\det g|} \left[- \frac12  f(\phi) R +
\frac{R^2}{6f_0^2} 
+ \sum_\phi \frac{(D_\mu\phi)(D^\mu\phi)}{2}- V_{\rm J}(\phi)\right] \eeq
In a purely dimension-less theory $V_{\rm J}$ is a sum of quartic interactions, and
$f = \sum_\phi \xi_\phi \phi^2$ is a sum of quadratic couplings.
The Planck mass can then arise from $f$, as scalar vacuum expectation values that break classical scale invariance (see e.g.~\cite{1502.01334,2205.06475}).
A renormalizable theory allows extra terms with positive mass dimension in $V_{\rm J}$ and $f$.

The $R^2$ term can be removed by 
adding the vanishing term $ - {(R+3f_0^2 \chi/2)^2}/{6f_0^2}$, where $\chi$ is an auxiliary field with no kinetic term.
Next, a Weyl rescaling $g_{\mu\nu}^{\rm E} = g_{\mu\nu} (f+\chi)/\bp^2$ allows to rewrite the action in the Einstein frame.
This generates an extra kinetic term showing that $R^2$ is equivalent to an extra scalar.
It is convenient to define such gravi-scalar as $z\equiv \sqrt{6(f+\chi)}$, such that the action in eq.\eq{S1} becomes~\cite{1502.01334}
\beq \label{eq:SEzphi}
S = \int d^4 x \sqrt{|\det g|} \left[- \frac{\bp^2}{2} R +
 \frac{6\bp^2}{z^2}\frac{(\partial_\mu z)^2+\sum_\phi (D_\mu \phi)^2 }{2} - V(\phi,z)\right]. \eeq
It has the form of eq.\eq{Lagen} with field space metric $K_{ab}=   \delta_{ab} 6 \bp^2/z^2$ that describes a warped space with 
constant  negative curvature ${\cal R}=- N_\phi(N_\phi+1)\bar /6\bp^2$, where $N_\phi$ is the total number of scalars $\phi$ apart from $z$.
We will compute using this basis where $z$ appears as a conformal factor in the field-space metric.
The Einstein-frame potential is
\beq \label{eq:VEgen}
V =  \left(\frac{6\bp^2}{z^2}\right)^2\left[V_{\rm J}(\phi) + \frac38 f_0^2 \left(f + \xi_z z^2\right)^2\right].
\eeq
with $\xi_z = -1/6$, corresponding to a conformal coupling.
So the gravi-scalar $z$, despite originating from the graviton,
 is just an ordinary scalar with
quartic self-coupling $\lambda_z \sim \xi_z^2 f_0^2 $ and mixed quartics
$\xi_z\xi_\phi f_0^2$ with other scalars $\phi$.
We will focus on the following scalars:
\begin{itemize}
\item  The `{\rm gravi-scalar}' $z$.

\item The {\rm Higgs} boson $h$  with
$V_{\rm J} \simeq \lambda_h h^4/4$ and $f = \xi_h h^2 + \bp^2$
has been considered in various cosmo-collider studies, finding that its large self-coupling $\lambda_h \sim 0.1$
allows cosmo-collider effects only in special corners of the parameter space.
So we will not consider the Higgs.

\item  A generic extra scalar singlet $s$.
Assuming $V_{\rm J} = \lambda_s(s) s^4/4$ and $f=\xi_s s^2$
gives a special motivated scalar $s$ dubbed `Planckion' because (unlike the Higgs) it dominantly induces the Planck mass as $\bp^2 = \xi_s s^2$. 
\end{itemize}
More in general, the Higgs $h$ together with an extra scalar singlet $s$ was studied in~\cite{1105.2284},
corresponding to $f = \bp^2 + \xi_h h^2 + \xi_s s^2$.

\begin{figure}[t]
\begin{center}
$$
\includegraphics[width=0.4\textwidth]{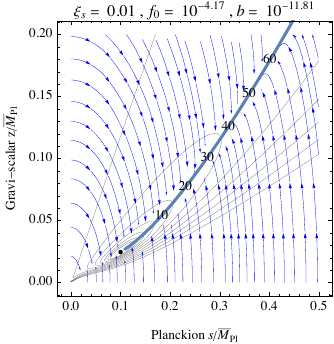}$$
\caption{\label{fig:trajectory}\em Sample of inflationary trajectory in the dimension-less theory of eq.\eq{SEzphi}
with a Planckion $s$ and the gravi-scalar $z$.
The blue arrows indicate field-space trajectories that converge in the inflationary trajectory with the indicated number of $e$-folds.
The thin lines are contour-levels of the potential.}
\end{center}
\end{figure}

\subsection{Computing the turn rate}
According to eq.\eq{etakappa}, the turn rate $\eta_N$ is written in terms of the curvature $\kappa$ of the inflationary trajectory.
Inflation proceeds along valleys of $V$ in eq.\eq{VEgen}, which are mildly curved in $z,\phi_a$ field coordinates
as illustrated in fig.\fig{trajectory}.
We approximate the inflationary trajectory as straight in field space,
and consider the effect of the warped kinetic metric in the theory of eq.\eq{SEzphi}.
We write the trajectory as $\phi_a = k_a z$ with generic slope defined by the $k_a$ parameters.
The curvature of this trajectory is
\beq\label{eq:kappastraight} \kappa=-\frac{1}{\bp\sqrt{6(1+1/\sum_a k_a^2)}}\le \frac{1}{\sqrt{6}\bp}.
\eeq
%
Next, we specialise this general expression to various well-studied models.
The form of the potential in eq.\eq{VEgen} implies that the inflaton tends to be the scalar combination with smallest quartic,
as it is the lightest scalar.
\begin{itemize}

\item {\em Dominant Starobinsky inflation}.
If $f_0^2$  is smaller than other scalar quartics, the inflaton dominantly is the Starobinsky gravi-scalar $z$.
Other scalars $\phi$ can be integrated out leaving $f\simeq \bp^2$ and $V \simeq 3\bp^4 f_0^2 (1-6\bp^2/z^2)^2/8$, describing Starobinsky inflation.
In this limit  $k_a \to 0$ and thereby  $\kappa\to 0$.
So the turn rate remains negligible in the quasi-single field regime where $z$ dominates inflation and
some extra scalar has Hubble mass.

\item {\em Dominant Higgs-like inflation}.
If $f_0^2$ is larger than other quartics, the gravi-scalar $z$ is heavy and can be integrated out as $z^2 \simeq 6 f$
leaving $V(\phi) = V_{\rm J} (\bp^2/f)^2$.
Assuming that the inflaton dominantly is $\phi_1=s$, the inflationary trajectory is $s \simeq k_s z$ with proportionality constant $1/k^2_s= 6\xi_s$.
According to eq.\eq{kappastraight}
the inflationary curvature is $\kappa \sim 1/\bp$ if $\xi_s\sim 1$.
This non-negligible turn rate persists in the quasi-single field regime where $s$ dominates inflation
and $z$ has Hubble-scale mass.

\end{itemize}
The spectral index $n_s$ agrees with data in both limits, as well as in the intermediate region~\cite{1502.01334}.
Motivated by these considerations we next compute $\fNL$ focusing on a model 
where the curvature saturates the bound of eq.\eq{kappastraight}.



\subsection{Inflation with gravi-scalar and Planckion}\label{adim}
As a first inflationary theory with two motivated scalars, we consider the Starobinsky gravi-scalar $z$
together with a `Planckion' scalar $s$ that dynamically generates the Planck mass.
The action of the theory is given by eq.\eq{SEzphi} with $f(s)=\xi_s s^2$.
 The non-vanishing field-space Christoffel symbols, Ricci tensor and curvature scalar are
\beq
\Gamma_{zz}^{z}=-\Gamma_{ss}^{z}=\Gamma_{zs}^{s}=\Gamma_{sz}^{s}=-\frac{1}{z},\qquad 
\mathcal{R}_{ab}= \frac{K_{ab}}{2}{\cal R}
,\qquad
\mathcal{R}=-\frac{1}{3\bp^2}.
\eeq
To achieve dynamical breaking of classical scale invariance at the Planck scale compatibly with QFT we assume
\beq \label{eq:lambda(s)}
V_{\rm J}(s) =\lambda_s(s)\frac{s^4}{4},\qquad \eeq
with running self-quartic $\lambda_s$
that reaches a minimal vanishing value at a scale $w$, $\lambda_s(w)=0$.
In this way the Einstein-frame potential $V$ of eq.\eq{VEgen}
has a minimum with vanishing value at $s=w$, as needed at inflation end.
We can neglect the RG running of $\xi_s$ and of other parameters.
This model provides a specific example of what we just discussed:
\begin{itemize}
\item In the limit 
$f_0^2 \ll \lambda_s$
the inflaton is the gravi-scalar $z$ while $s\simeq w$ sits around its minimum,
giving Starobinsky inflation with $f_0\approx 10^{-5}$, small $r$ and negligible turn rate.

\item In the opposite limit 
$\lambda_s \ll f_0^2$ one gets Planckion inflation along the trajectory $z^2 \simeq 6 \xi_s s^2$
such that the inflationary potential and $V_{NN}$ are
\beq V \simeq \frac{\lambda_s(s)\bp^4}{4\xi_s^2},\qquad \frac{V_{NN}}{H^2}\simeq \frac{6 f_0^2 \xi_s^2 }{\lambda_s(s)}(1 +6 \xi_s).\eeq
The turn rate is $\eta_N \sim \sqrt{\epsilon_H}$ and the dominant cubics are
\beq \label{eq:VNNetc}
\frac{V_{NNN}}{H} \simeq \frac{f_0^2\xi_s}{\sqrt{\lambda_s(s)}}3 \sqrt{2}(1+3\xi_s)\sqrt{1+6\xi_s},\qquad
\frac{V_{NTT}}{H} \simeq  \frac{f_0^2\xi_s}{\sqrt{2\lambda_s(s)}}\sqrt{1+6\xi_s}.
\eeq
\end{itemize}
Cosmo-collider signals arise in the quasi-single field regime where
the Planckion $s$ is the inflaton, and the gravi-scalar $z$ has
Hubble mass, $V_{NN}/H^2 \sim 1$ at ${\cal N}\approx 60$ $e$-folds of inflation.
Eq.\eq{VNNetc} shows that potential cubics get small in such regime, 
\beq V_{NTT,NNN}/H \sim f_0 \sim H/\bp.\eeq
Extra contributions to non-derivative cubics proportional to the Riemann tensor
(see eq.\eq{L3scalflat} and\eq{L3scalcomoving}) are suppressed by extra slow-roll parameters
compared to the scalar potential contribution.
The non-Gaussianity $\fNL$ of eq.\eq{fNLcc} is dominated by the single-isocuvaton exchange diagram, and is estimated as
\beq \fNL \sim   \frac{\eta_N}{\sqrt{P_\zeta}} \frac{V_{NTT}}{H}\sim
\eta_N \sqrt{\epsilon_H}
 \sim \epsilon_H  \eeq
as $P_\zeta \simeq (H/\bp)^2/8\pi^2\epsilon_H$.
The final result $\fNL\sim\epsilon_H$ does not depend on the specific form $\lambda_s(s)$ of the Planckion potential that determines $\epsilon_H$.
This cosmo-collider effect is parametrically similar to the minimal gravitational contribution to $\fNL\sim \epsilon_H$.
The shape dependence is different, as cosmo-collider effects appear in the squeezed limit.

All the above model-specific results agree with the model-independent considerations presented around eq.~(\ref{sys:V'''}).
We there identified one more possibly relevant contribution to $\fNL \propto V_{NNN}$, from triple-isocurvaton exchange.
This extra contribution is negligible in the present model, because it predicts a small
$V_{NNN}\sim V_{NTT}$ as in eq.\eq{VNNetc}.

\medskip
%


Finally, a numerical computation is needed to precisely compute the cosmo-collider $\fNL$ signal.
To proceed, we specify the RG running of $\lambda_s$ approximating it at leading order in couplings to be
\beq\label{eq:lambdasrun}
 \lambda_s(s) \simeq  \frac{b}{8}\ln^2 \frac{s^2}{w^2}\eeq
where the $\beta$-function coefficient $b$ is naturally given by $b= g^4/(4\pi)^4$ where $g$ is some unspecified dimension-less coupling.


\begin{figure}[t]
\begin{center}
$$\includegraphics[width=0.4\textwidth]{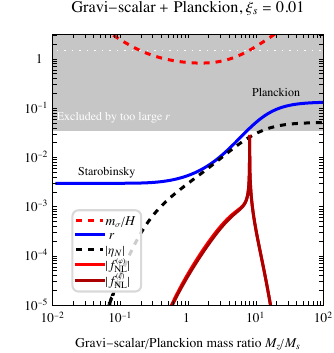}\qquad
\includegraphics[width=0.4\textwidth]{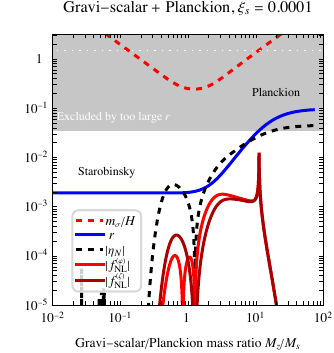}$$
\caption{\label{fig:Isz}\em Predictions of inflation with two scalars: gravi-scalar $z$ and Planckion $s$.
The blue curves show the tensor/scalar ratio $r$ (the gray region is excluded);
the black dashed curves show the turn rate $\eta_N$;
the nearly-overlapping red curves show the value of $\fNL$, computed in $\varphi$-gauge (lighter red) and in $\zeta$-gauge (darker red).
}
\end{center}
\end{figure}

In general, inflation models with multiple scalars can have different inflationary trajectories depending on the starting point.
In this model, as typical of dimension-less models, inflation proceeds along the bottom of the potential energy valley.
The example in fig.\fig{trajectory} show how different initial conditions converge towards one common inflationary trajectory,
mildly curved in field space.
So there is no dependence on initial conditions, and the model has four free parameters: $f_0,b, \xi_s, w$.
The latter is fixed by $\xi_s w^2 = \bp^2$, and one combination of $f_0$ and $b$ is fixed by demanding $P_\zeta \approx 2.1~10^{-9}$.
This leaves two free parameters, that we choose to be $\xi_s$ and $M_z/M_s$.
Here 
\beq M_z = f_0 \bp/\sqrt{2},\qquad M_s = \bp\sqrt{4b/\xi_s(1+6\xi_s)}\eeq 
are the gravi-scalar mass and the Planckion mass,
both computed after inflation and in the limit $M_z\gg M_s$ or $M_z\ll M_s$ where their mass mixing is negligible.
Cosmo-collider signals arise in the intermediate regime where $s$ has a mass comparable to the Hubble scale.

Fig.\fig{Isz} shows inflationary prediction for fixed values of $\xi_s $ as function of $M_z/M_s$. 
The isocurvaton mass (upper red dashed curve) crosses the optimal cosmo-collider value $m_\sigma/H \approx 3/2$
twice: 
for $z$-inflaton and $s$-isocurvaton (giving tiny $\fNL$);
for $s$-inflaton and $z$-isocurvaton (giving $\fNL$ at gravitational floor level).
In the second case the tensor/scalar ratio $r$ is just below current bounds, maximising $\fNL$.
This coincidence is a prediction of the model and was not imposed by tuning its parameters.\footnote{Inflation along $s$, rewritten in terms of a canonically normalised $s_{\rm can}$, is  equivalent to quadratic inflation, that predicts a too large tensor/scalar ratio $r$.
It becomes compatible with bounds when $z$ is light enough to give cosmo-collider effects.
In more general models, an extra cubic term in eq.\eq{lambdasrun} would modify the value of $r$~\cite{1502.01334}. 
However, a significant cubic signals that higher-order terms in the Taylor expansion cannot be neglected.
Full theories with the desired RG running have been studied in~\cite{2109.10367}.}
At the cosmo-collider point the turn rate is $\eta_N \approx 0.025$ inducing $\fNL$ at the few per-mille level.
Computations of $\fNL$ performed in the $\varphi$-gauge and $\zeta$-gauge agree, up to neglected sub-leading terms.





\begin{figure}[t]
\begin{center}
$$\includegraphics[width=0.4\textwidth]{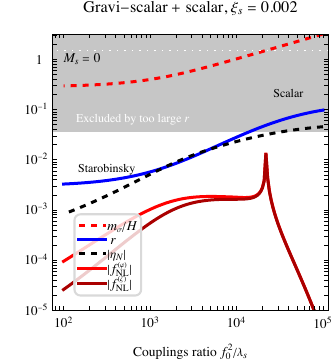}\qquad
\includegraphics[width=0.4\textwidth]{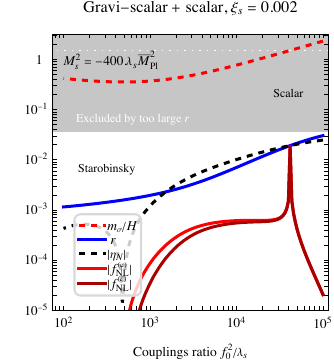}$$
\caption{\label{fig:Isz2}\em Predictions of inflation with gravi-scalar $z$ and generic scalar $s$.
Curves as in fig.\fig{Isz}.
{\bfseries Left panel}: a Higgs-like scalar with a constant quartic $\lambda_s$.
{\bfseries Right panel}: a scalar with a constant quartic $\lambda_s$ and a mass term.}
\end{center}
\end{figure}

\subsection{Inflation with gravi-scalar and a generic scalar}\label{dim}
The previous model can be generalized dropping the assumption that the scalar $s$ dynamically generates the Planck scale.
The scalar $s$ could dynamically generate sub-Planckian scales, such as the unification scale, the right-handed neutrino mass scales, the weak scale...
This kind of  low-scale behaviour is largely irrelevant during inflation, when the inflaton $s$ is Planckian.
One can then quite generically approximate the part of the $s$ action relevant for inflation as eq.\eq{SEzphi} with
\beq V_{\rm J} (s) = \frac{M_s^2}{2} s^2 + \frac{V_{sss}}{6} s^3+\frac{\lambda_s}{4} s^4,\qquad f = \bp^2 + \xi_s s^2\eeq
where dimension-full terms break classical scale invariance.

We first consider a scalar with $M_s, V_{sss} \simeq 0$ and constant quartic $\lambda_s$.
Such a scalar is sometimes dubbed `Higgs', but improperly:
cosmo-collider signals arise at $m_\sigma\sim H$, when $\lambda_s \sim f_0^2$ is much smaller than the typical Higgs quartic.
More precisely, the iso-curvaton mass at inflationary values $z\gg \bp/\sqrt{\xi_s}$ is 
\beq \frac{m_\sigma^2}{H^2} \simeq 24\xi_s \left[1+ \frac{ f_0^2 \xi_s}{4\lambda_s}(1+6\xi_s) \right]- \eta_N^2\eeq
showing that a mildly small $\xi_s$ is also needed to have $m_\sigma\simeq 3H/2$.
In this model inflationary trajectories again have Planckian curvature $|\kappa|\sim 1/\bp$,
giving a turn rate again suppressed by the slow-roll parameter as in eq.\eq{etakappa},
and the typical small cosmo-collider signal of this class of models, $\fNL \sim 0.01$.
This is confirmed by the numerical result in the left panel of  fig.\fig{Isz2}.\footnote{This model was recently studied in~\cite{2309.10841},
claiming a large $\fNL$ produced by a large turn rate, $|\eta_N|\sim 1$ or even larger.
We disagree with the claim, due to an invalid approximation used in~\cite{2309.10841} to compute $\eta_N$, as we now discuss.
Eq.\eq{eqTN} implies $\eta_N = V_N/H^2 \phi'_0$ which, in slow roll approximation, becomes $\eta_N \simeq 3 V_N/V_T$.
Since both the numerator and denominator vanish in slow-roll approximation, this approximation can be used only if the trajectory is computed exactly.
An incorrect $\eta_N \sim 1$ is obtained using instead the approximated slow-roll trajectory assumed in~\cite{2309.10841}.}

\smallskip

The right panel of  fig.\fig{Isz2} adds a mass term $M_s^2$. 
As expected, a Planckian mass term is needed to have effects.
Even so, only order unity factors change.
A similar conclusion holds for cubic terms.

A scalar $s$ with $\xi_s=0$ and a potential minimum at $s\neq 0$ was studied in~\cite{2404.05031}, 
finding a tiny $\fNL\sim 10^{-8}$ assuming gravi-scalar-dominated inflation.
Consistently with our previous discussion, a larger $\fNL$ needs that $s$ contributes significantly to inflation.

\section{Big cosmo-collider signals from multi-scalar inflation?}\label{rot}
Based on our previous considerations, we can now discuss if/which
multi-scalar inflation models can lead to larger cosmo-collider signals.
This needs a `large' turn rate not suppressed by slow-roll parameters,
such as $\eta_N\sim 1$.\footnote{Our perturbative expansion for $\fNL$ breaks down at larger $|\eta_N| \gg 1$,
a regime dubbed `rapid-turn inflation', see e.g.~\cite{1510.01281,1804.11279,1805.12563,1902.03221,2111.00989,2210.00031,Kolb:2022eyn,2405.11595}.}
Eq.\eq{etakappa} implies that $\eta_N$ can be increased by increasing  $\epsilon_H$ and/or the curvature $\kappa$ of the inflationary trajectory.

Increasing the slow-roll parameter $\epsilon_H$  is difficult because inflation needs $\epsilon_H \ll  1$ and because $\epsilon_H$ is subject to two constraints:
\begin{enumerate}
\item from the upper bound on the tensor-to-scalar ratio $r\approx 16 \epsilon_H$;
\item from its contribution to the measured spectral index $n_s = 1- 2\epsilon_H + \cdots$.
\end{enumerate}
Circumventing both constraints at the same time seems unappealing.
For example, 4-derivative gravity predicts a spin-2 ghost that
makes $r < 16 \epsilon_H$ when lighter than the Hubble scale~\cite{2202.00684}.
This would allow to bypass the first issue.
To bypass the second issue one might break the usual relation between $\epsilon_H$ and the number ${\cal N}\approx 60$ of $e$-folds at CMB scales
by abruptly ending inflation adding an extra scalar to implement the water-fall mechanism.

We do not pursue these possibilities, and explore increasing $\kappa$.
Section~\ref{globalturn} discusses the possibility that many turns $\Delta\theta\equiv  \int |\eta_N| d{\cal N}\gg 2\pi$ occured during inflation,
at a roughly constant rate.
Section~\ref{localturn} discusses the possibility that a limited turning $\Delta \theta \sim 1$ occured,
but concentrated around the CMB scale at ${\cal N}\approx 60$ $e$-folds before the end of inflation.


\begin{figure}[t]
\begin{center}
$$\includegraphics[width=0.4\textwidth]{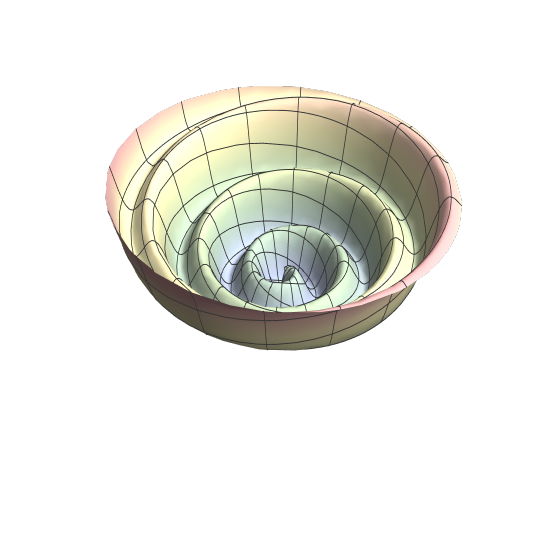}\quad
\includegraphics[width=0.4\textwidth]{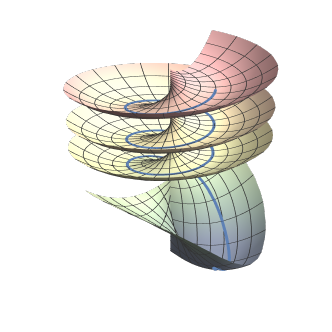}$$
\vspace{-12ex}
\caption{\label{fig:mono}\em Qualitative examples of inflationary potentials (as function of two scalars)
that allow multiple turns with negative curvature $\kappa$.}
\end{center}
\end{figure}

\subsection{Larger turn rate during all inflation}\label{globalturn}
According to eq.\eq{etakappa}, a larger turn rate arises if the 
curvature $\kappa$ of the inflationary trajectory is produced by sub-Planckian physics
and thereby larger than $|\kappa|\sim 1/\bp$. 
A large and roughly constant turn rate implies many turns $\Delta \theta = \int |\eta_N| d{\cal N}$
during the ${\cal N}\gtrsim 60$ $e$-folds of inflation,
implying special geometries of the inflationary trajectory.

Section~\ref{for} started from the action written in generic basis in scalar field space
and identified two useful combinations: the inflaton $\phi$ and the orthogonal isocurvaton $\sigma$.
It is now convenient to promote $\phi$ and $\sigma$ to a coordinate system for  
the scalar field space (at least locally) and write the action  these convenient coordinates:
\beq \Lag = K_{\sigma\sigma} \frac{(\partial_\mu \sigma)^2}{2} + K_{\phi\phi} \frac{(\partial_\mu \phi)^2}{2}- V (\phi,\sigma) .
\eeq
The kinetic coefficients $K_{\sigma\sigma}$ and $K_{\phi\phi}$ are two functions of $\phi,\sigma$.
Without loss of generality one could make $\sigma$ canonically normalized around the inflationary trajectory, setting $K_{\sigma\sigma}=1$.
The potential $V$ must then have a form consistent with $\phi$ being the inflaton, such as
\beq
V (\phi,\sigma) =  \delta V(\phi) + \frac{m^2_\sigma}{2} (\sigma-v_\sigma)^2 + \frac{V_{\sigma\sigma\sigma}}{6}(\sigma - v_\sigma)^3 + \cdots
\eeq
where $\delta V$ is any inflationary potential.
The curvature of the inflationary trajectory at roughly fixed $\sigma \approx v_\sigma$ is\footnote{The centrifugal force
makes $\sigma> v_\sigma$ and becomes significant when the turn rate is large.
Special potentials that keep $\sigma$ constant can be written, but have undesired features around inflation end~\cite{1901.03657}.
}
\beq \kappa = -\frac{1}{2\sqrt{K_{\sigma\sigma}}} \frac{d}{d\sigma}\ln K_{\phi\phi},\eeq
not affected by a multiplicative constant in $K_{\phi\phi}$. 
This is related to the curvature of field space as
${\cal R}=2{\kappa^2}\left[1 -2  {K_{\phi\phi}K''_{\phi\phi}}/{K_{\phi\phi}'^2}\right]$.
Large negative or positive curvature $\kappa$ can be obtained as follows: 
\begin{itemize}
\item $K_{\phi\phi}=\sigma^2$ means ${\cal R}=0$: the field space is a flat Euclidean space in polar coordinates, 
and the inflationary trajectory is a circumference with radius $r=\sigma$, such that $\kappa=-1/r$.
However, a U(1) or some other global symmetry broken at $\sigma\ll \bp$ is not enough, because
a large turn rate implies multiple turns during all inflation.
Potentials such as those in fig.\fig{mono} are needed:
either a long valley that spirals towards a central point as illustrated in the left panel of fig.\fig{mono},
or a potential $\delta V$ not periodic under rotations $\phi\to \phi + 2\pi f$
as illustrated in the right panel of fig.\fig{mono}~\cite{Chen:2009zp,0909.0496,Achucarro:2012yr,Kolb:2022eyn}.\footnote{This  
feature can arise in theories where two axions have vastly different periodicities $f_1\gg f_2$~\cite{0912.1341}.
These models however give a constant $K_{\phi\phi}$ and need
vastly different charges, that can maybe justified as in~\cite{1511.01827}.
Another possibility is a field $\phi$ that describes, in the low energy 4-dimensional effective field theory,
 the amount of wrapping of a brane along some compactified extra dimension.
The energy $\delta V$ keeps changing during multiple wrappings~\cite{0803.3085},
similarly to the magnetic energy of a solenoid.}
This allows to increase the single-isocurvaton exchange contribution to $\fNL\sim \eta_N^2$.
Furthermore, the triple-isocurvaton contribution becomes significant if, additionally, $V_{\sigma\sigma\sigma}$ is large enough~\cite{Chen:2009zp}.

\item $K_{\phi\phi}=\mu^2_\sigma/\sigma^2$ means $\kappa=1/r$ with radius $r=\sigma$ in the basis $K_{\sigma\sigma}=1$.
 The field space has constant negative curvature ${\cal R} =-4 \kappa^2$.
This was realized in the models of section~\ref{scal}, where $K_{ss}=K_{zz}=6\bp^2/z^2$.
In the $f_0^2\gg \lambda_s$ and $\xi_s\ll 1$ regime
the inflaton was $\phi \simeq s$ and the isocurvaton was the  gravi-scalar $\sigma\simeq z$,
realising the situation here discussed.
Increasing the turn rate requires adding effective operators suppressed by a sub-Planckian scale such that 
$\mu_\sigma\ll \bp$.\footnote{A possible motivation for modifying kinetic terms is gravity in the Palatini formalism, where the connection $\Gamma^\mu_{\nu\rho}$ is considered as a field independent of the metric.
In two-derivative Einstein gravity $\Gamma$ is a non-dynamical field:
its non-derivative field equations imply the standard expression for $\Gamma$ in terms of derivatives of the metric.
In the presence of scalars with non-minimal coupling, $f(\phi)R$,
Einstein and Palatini gravity differ by specific extra Planck-suppressed operators~\cite{2002.07105}, 
and thereby describe in different ways the same general physics.
The Palatini formalism implies extra physical degrees of freedom in theories of 4-derivative gravity, as they make $\Gamma$ dynamical. 
A field redefinition shows that Palatini is equivalent to standard gravity plus extra higher-spin fields related to $\Gamma$~\cite{2006.01163}.
Higher-spin fields are in general problematic, as they contain components with negative kinetic energy
(in addition to the spin-2 ghost predicted by 4-derivative gravity).
If the action only contains the $R^2$ term, the Palatini formulation reduces to a scalar with a special combination of higher derivatives~\cite{1810.05536}.
}
Inflationary models with hyperbolic field space have  been studied in~\cite{1510.01281,1707.05125,1804.11279,1902.03221,2111.00989,2304.14260}.
\end{itemize}
A new feature appears in theories with 3 or more scalars: the inflationary curvature can have a torsion, in addition to a turn rate~\cite{2011.05930,2304.00065}.

\smallskip





\begin{figure}[t]
\begin{center}
$$\includegraphics[width=\textwidth]{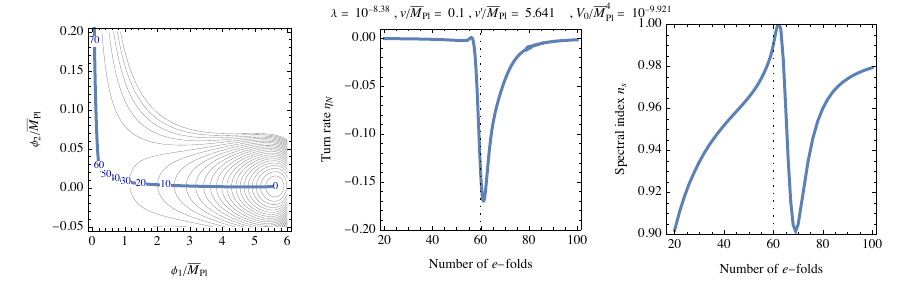}$$
\caption{\label{fig:Lvalley}\em Inflation in the model of eq.\eq{Lvalley},
along a valley that turns at ${\cal N}\approx 60$, generating a temporarly large turn rate.
The left panel shows the inflationary trajectory at various ${\cal N}$ together with contour-levels of the potential;
the middle panel shows the turn rate peaked at ${\cal N}\approx 60$;
the right panel shows the consequent feature in the spectral index.
}
\end{center}
\end{figure}

\subsection{Larger turn rate around the CMB scale?}\label{localturn}
We here consider the possibility of a large turn rate present only when, during inflation, the current cosmological scales
exited the horizon. The power spectrum can be computed from the
Mukhanov-Sasaki equations for the time evolution of scalar perturbations.
These arise by performing functional derivatives of the $\varphi$-gauge
quadratic action in eq.\eq{S2varphigauge} or of the $\zeta$-gauge action in eq.\eq{S2zetagauge}~\cite{Achucarro:2012yr}.
In slow-roll approximation:
\begin{eqnsystem}{sys:MS}
\zeta''_k + 3 \zeta'_k + \frac{k^2}{H^2} e^{-2{\cal N}}\zeta_k &=& - 2 (3\eta_N \sigma_k +\eta_N \sigma'_k + \eta'_N \sigma_k ),\\
 \sigma''_k + 3 \sigma'_k +\frac{m_\sigma^2 + k^2 e^{-2{\cal N}} }{H^2} \zeta_k  &=& 2 \eta_N  \zeta_k.
\end{eqnsystem}
These equations show that a turn  rate $\eta_N$ transfers $\sigma_k$ fluctuations into $\zeta_k$ curvature fluctuations, enhancing or suppressing them.
In particular, a sharp turn with total angle $\Delta\theta$
enhances $P_\zeta(k) $ by $\sim e^{\Delta\theta }$ for $k$ values that experience the turn around horizon exit,
also leading to oscillatory features~\cite{2004.06106,2004.08369,2012.02761,2111.14664,2112.06903,2203.15605}.
A signal in $\fNL$ is thereby constrained by the observed $P_\zeta$ (see also~\cite{2210.07028}).

\medskip

As a concrete example, one can write inflation potentials where inflation happens along a nearly-flat valley that, at some point, exhibits a turn.
Considering two scalars $\phi_1,\phi_2$ 
with canonical kinetic term, the potential
\beq \label{eq:Lvalley}
V= \frac{\lambda}{4}(\phi_1 \phi_2 - v^2)^2 + \delta  V(\phi_1,\phi_2)\eeq
describes a valley with $\phi_2\gg\phi_1 $ that turns by $\Delta\theta=\pi/2$ into a valley with $\phi_1\gg \phi_2 $
around the point $\phi_1 = \phi_2 = v$.
These valleys are flat if $\delta V=0$;
this extra term can be any smooth potential that adds a small slope along the valleys.
For example we consider $\delta V$ to be of Starobinsky form $\delta V = V_0 (1 - e^{-c (\phi_1 + \phi_2-v')/\bp})^2$.
The free parameter $V_0$ is chosen to reproduce the observed $P_\zeta$;
the quartic coupling $\lambda$ is chosen such that $m_\sigma \sim H$;
$v'$ is chosen such that the turn happens at ${\cal N}\approx 60$;
a sub-Planckian $v \sim 0.1\bp$ is chosen to concentrate the $\Delta\theta$ turn in a few $e$-folds of inflation.
We set $c=\sqrt{2/3}$.

Fig.\fig{Lvalley} shows the numerical solution.
A sharp turn (left panel of fig.\fig{Lvalley}) happens at ${\cal N}=60$ 
with $|\eta_N|\approx 0.14$ (middle panel of fig.\fig{Lvalley})
while $m_\sigma/H \approx 3/2$.
As expected, the resulting non-Gaussianity $\fNL\approx 0.3$
is accompanied by an effect in the power spectrum (right panel of fig.\fig{Lvalley}):
its spectral index oscillates around $n_s\approx 0.96$ by a few $\%$.
Having assumed $\delta V$ with Starobinsky form, the tensor to scalar ratio is small, $r\approx 3~10^{-3}$.
Different choices for $\delta V$ that lead to a larger $r$ also lead to a larger $\fNL$.

\smallskip

A qualitatively different possibility is a potential with two different valleys that merge, forcing the inflaton to turn.
Considering again two scalars $\phi_1,\phi_2$ with canonical kinetic term, a possible potential is
\beq \label{eq:Xvalley}
V= \frac{m^2}{2} \frac{\phi_1^2 \phi_2^2}{\phi_1^2+\phi_2^2}  + \delta V(\phi_1,\phi_2).\eeq
Keeping the same $\delta V$, a numerical solution shows that 
this merging structure leads to a much larger enhancement and local features in $P_\zeta$.
Models of this type have been indeed considered for generating primordial black holes at small scales~\cite{2012.03705,2005.02895}:
for the same reason they seem unsuitable for generating non-Gaussianities at CMB scales.

In the context of $R^2$ gravity, \cite{1712.09896} considered gravi-scalar inflation that turns into $s$ inflation,
thanks to a potential of the form $V(s) = V_0  - M_s^2 s^2/2$,
finding that the turn is accompanied by a large feature in $P_\zeta(k)$.

\section{Conclusions}\label{concl}
Non-Gaussianity signals due to particles with Hubble scale masses
during inflation are known as `cosmo-collider',  
because they can be seen as particle production from inflaton scatterings.
These signals are however small in most models of inflation,
where the near flatness of the inflationary potential corresponds to the inflaton being a nearly free particle.
In collider language, the cosmo-collider has high luminosity but cross sections are small.
Without giving up on the possibility that primordial inhomogeneities are inflaton perturbations,
we explored the possibly that large enough {\em tree-level} effects that can arise when the extra particle is a {\em scalar}.

\medskip

In section~\ref{for} we critically summarized the non-trivial formalism that allows to compute such cosmo-collider effects.
Tree-level signals arise when the inflationary trajectory is curved in field space,
such that the turn rate $\eta_N $ transmits non-Gaussianities from extra `isocurvaton' scalars $\sigma$ into the inflaton.
We identified two dominant contributions to the $\fNL$ parameter that controls non-Gaussianities:
\begin{itemize}
\item  single-isocurvaton exchange gives $\fNL \sim \eta_N^2$.

\item  triple-isocurvaton exchange gives a contribution 
$ \fNL \sim  \eta_N^3 \sqrt{\epsilon_H}  V_{NNN}/(H^2/\bp)$ 
suppressed by an extra power of the turn rate, and possibly enhanced by the isocurvaton cubic coupling $V_{NNN}$
that can be large.
\end{itemize}
So a large $\fNL$ needs a large turn rate.
We emphasized the intuitive relation
\beq \hbox{(turn rate $\eta_N$)} = \hbox{(curvature $\kappa$ of the inflationary trajectory)} \cdot \hbox{(inflaton velocity)}\eeq
where the inflaton velocity $\sim \sqrt{\epsilon_H}\bp$ is slow-roll suppressed.
This implies that a large class of inflation models where Planckian physics induces a Planck-suppressed curvature $\kappa\sim 1/\bp$
predict cosmo-collider signals $\fNL \sim \epsilon_H$, at the same level of the gravitational floor but with a distinctive shape.
We precisely computed various examples of such models, finding that $\fNL \sim \epsilon_H$
indeed arises (compatibly with other current data) when the iso-curvaton is the gravi-scalar predicted by $R^2$ gravity.
In particular, dimension-less theories naturally lead to a Hubble-scale isocurvaton.
Results are shown in fig.\fig{Isz} and\fig{Isz2}.

\smallskip

However, $\fNL \sim \epsilon_H$ is below the sensitivity of the next round of observations.
We thereby explored various ways in which inflaton models might allow for larger cosmo-collider signals.
The main possibility compatible with current data is that sub-Planckian physics
induces a larger $|\kappa|\gg 1/\bp$ and roughly constant curvature of the inflationary trajectory.
As discussed in section~\ref{globalturn} this implies multiple turns during the $\gtrsim 60$ $e$-folds of inflation, and thereby special inflationary models,
such as a large positive or negative curvature in field space.


\small

\paragraph{Acknowledgements}
We thank Chao Chen, Yohei Ema, Lucas Pinol, Sebastien Renaux-Petel, Fumiya Sano, Xi Tong, Masahide Yamaguchi, Ivonne Zavala, Yuhang Zhu, and Zhong-Zhi Xianyu for discussion.  S.A. is supported by IBS under the project IBS-R018-D1.

\begin{appendix}

\section{Calculation of bispectrum}\label{calculation of bispectrum}

\subsection{Mode functions and propagators}
We quantise fluctuations of the nearly-massless inflaton $\varphi$ scalar as
\begin{align}
\varphi(\tau,\mb{x})&= \int \frac{\mathrm{d}^3 \mb{k}}{(2 \pi)^3}\left(\varphi_k(\tau) a_{\boldsymbol{k}}+
\varphi_k^*(\tau) a_{-\boldsymbol{k}}^{\dagger}\right)e^{i\boldsymbol{k} \cdot \boldsymbol{x}}, \label{Q_phi}
\end{align}
where $a_{\boldsymbol{k}}~(a_{\boldsymbol{k}}^{\dagger})$ is a creation (annihilation) operator satisfying the usual commutation relations
and $\tau$ is the proper time.
Similar expressions hold for the isocurvaton $\sigma$.
The mode functions are
\beq
 \varphi_k=\frac{H}{\sqrt{2 k^3}}(1+i k \tau) e^{-i k \tau}, \qquad
\sigma_k=-i  \frac{\sqrt{\pi}}{2} e^{i\left(\nu+\frac{1}{2}\right) \frac{\pi}{2}}H(-\tau)^{3 / 2} H_{\nu}^{(1)}(-k \tau),   
 \label{sol_chi} 
\eeq
where $H_\nu^{(1)}$ is the Hankel function of the first kind and 
$\nu$, given in eq.\eq{S_func},
is real for $m_\sigma<3H/2$ and purely imaginary for $m_\sigma>3H/2$. 

We compute the bispectrum using the Schwinger-Keldysh (SK) diagrammatic rules in~\cite{Chen:2017ryl}. 
The bulk-to-boundary propagators for $\varphi$ are
\begin{align}
G_{a}\left(k; \tau\right) = \frac{H^2}{2 k^3}(1-i a k \tau) e^{i a k \tau}, \label{G_a}   
\end{align}
where $a=\pm$ is the SK index. 
The bulk-to-bulk propagators $D_{ab}$ for $\sigma$ are 
\begin{align}
&D_{-+}\left(k ; \tau, \tau^{\prime}\right)=\sigma_k\left(\tau\right) \sigma_k^{*}\left(\tau^{\prime}\right)=\frac{H^2\pi e^{-\pi \operatorname{Im} \nu}}{4}  \left(\tau \tau^{\prime}\right)^{3/ 2} \mathrm{H}_\nu^{(1)}\left(-k \tau\right) \mathrm{H}_{\nu^*}^{(2)}\left(-k \tau^{\prime}\right),\\
& D_{+-}\left(k ; \tau, \tau^{\prime}\right)=\left(D_{-+}\left(k ; \tau, \tau^{\prime}\right)\right)^*,\\
&D_{ \pm \pm}\left(k ; \tau, \tau^{\prime}\right)=D_{\mp\pm}\left(k ; \tau, \tau^{\prime}\right) \theta\left(\tau-\tau^{\prime}\right)+D_{\pm\mp}\left(k ; \tau, \tau^{\prime}\right) \theta\left(\tau^{\prime}-\tau\right),\label{D_ppmm}
\end{align}
where $\theta$ is the unit step function. 
It is useful to combine the above two propagators defining a mixed propagator~\cite{Chen:2017ryl}, 
\begin{align}
\mathcal{G}_a(k;\tau) \equiv  \frac{2i\eta_N}{H^2} \int_{-\infty}^0 \frac{d \tau^{\prime}}{\left(-\tau^{\prime}\right)^3} \sum_{b= \pm}b\ D_{a b}\left(k ; \tau, \tau^{\prime}\right) \partial_{\tau^{\prime}} G_b\left(k ; \tau^{\prime}\right).   \label{mix}
\end{align}
By substituting the explicit form of the mode functions, the mixed propagators are expressed as
\begin{align}
\mathcal{G}_a(k;\tau)=   \frac{\pi \eta_N H^2 }{4 k^3} \mathcal{I}_{a}(z),  
\end{align}
where $z\equiv -k\tau$ and
\begin{align}
\nonumber \mathcal{I}_{ a}(z)=&\ e^{-\pi \operatorname{Im} \nu} z^{\frac{3}{2}}\left\{2 \operatorname{Im}\left[H_{\nu}^{(1)}(z) \int_0^{\infty} \frac{d z^{\prime}}{\sqrt{z^{\prime}}} H_{\nu^*}^{(2)}\left(z^{\prime}\right) e^{-i z^{\prime}}\right]\right.\\
&+i H_{\nu}^{(1)}(z) \int_0^z \frac{d z^{\prime}}{\sqrt{z^{\prime}}} H_{\nu^*}^{(2)}\left(z^{\prime}\right) e^{- i a z^{\prime}}\left.-i H_{\nu^*}^{(2)}(z) \int_0^z \frac{d z^{\prime}}{\sqrt{z^{\prime}}} H_{\nu}^{(1)}\left(z^{\prime}\right) e^{- i az^{\prime}}\right\}.\label{I_func}
\end{align}
The $z$-integral can be performed analytically, see~\cite{Chen:2017ryl} for the explicit expression. 
For our purposes, the behavior in the $z\rightarrow 0$ limit is useful:
\begin{align}
\mathcal{I}_{ a}(z) \stackrel{z \rightarrow 0}{\longrightarrow}-\frac{2}{\sqrt{\pi}}\left[\frac{2^{\nu} z^{3 / 2-\nu} \Gamma(\nu)}{\cos\left(\frac{\pi \nu}{2}\right)+\sin \left(\frac{\pi \nu}{2}\right)}+(\nu \rightarrow-\nu)\right].    \label{asymp}
\end{align}

\subsection{Bispectrum}
We focus on the three tree level diagrams in fig.\fig{Feyn}, build viewing as perturbations the mixing term in eq.\eq{Lvarphisigma}
and the cubic couplings in eq\eq{Vabc}.
The scale factor is $a=(-H\tau)^{-1}$ in de Sitter space.
Following the SK diagrammatic rules in~\cite{Chen:2017ryl}, the contribution to the  bispectrum from the first term in single $\sigma$-exchange diagram in fig.\fig{Feyn} is 
\begin{align}\label{eq:NTT}
\left\langle\varphi_{\mathbf{k}_1} \varphi_{\mathbf{k}_{\mathbf{2}}} \varphi_{\mathbf{k}_{\mathbf{3}}} \right\rangle^{\prime}_{NTT}   
= \frac{2\eta_N V_{NTT}}{H^6} \sum_{a, b= \pm}(a b) \int_{-\infty}^0 \frac{d \tau_1}{\left(-\tau_1\right)^4} \frac{d \tau_2}{\left(-\tau_2\right)^3} 
\bigg[G_a\left(k_1 ; \tau_1\right) G_a\left(k_2; \tau_1\right)\\
  D_{a b}\left(k_3 ; \tau_1, \tau_2\right) \partial_{\tau_2}G_b\left(k_3 ; \tau_2\right)+
 (k_1 \leftrightarrow k_3) + (k_2 \leftrightarrow k_3)\bigg]\nonumber
\end{align}
where the prime on $\left\langle\varphi^3\right\rangle$ denotes that a momentum conservation factor $(2\pi)^3\delta^{(3)}({\mb{k}}_1 +{\mb{k}}_2+{\mb{k}}_3)$ is extracted.
Inserting the mixed propagator in eq.~\eqref{mix}, it becomes
\begin{align}
\left\langle\varphi_{\mathbf{k}_1} \varphi_{\mathbf{k}_{\mathbf{2}}} \varphi_{\mathbf{k}_{\mathbf{3}}} \right\rangle^{\prime}_{NTT} 
&=\frac{-iV_{N TT}}{ H^4}\sum_{a= \pm}a\ \int_{-\infty}^0 \frac{d \tau_1}{\left(-\tau_1\right)^4} G_a\left(k_1; \tau_1\right) G_a\left(k_2 ; \tau_1\right) \mathcal{G}_a\left(k_3;\tau_1\right)+P\\
&=\frac{\pi\eta_NV_{NTT} H^2}{16i k_2^3k_3^3}\sum_a a \int_0^{\infty} \frac{d z}{z^4}(1+i a z)\left(1+i a \frac{k_2}{k_1} z\right) e^{-i a\left(1+\frac{k_2}{k_1}\right) z}\mathcal{I}_a\left(\frac{k_3}{k_1} z\right)+P .\label{B_a}
\end{align}
where $P$ denotes two permutations.\footnote{The analogous $\zeta$-gauge diagram with derivative cubic interactions
leading to eq.\eq{fNLzeta}
is computed replacing 
$ G_a\left(k_1 ; \tau_1\right) G_a\left(k_2; \tau_1\right)$ in eq.\eq{NTT}
with  
$\partial_{\tau_1}G_a\left(k_1 ; \tau_1\right) \partial_{\tau_1} G_a\left(k_2; \tau_1\right)
+\mb{k}_1\cdot\mb{k_2}G_a\left(k_1 ; \tau_1\right) G_a\left(k_2; \tau_1\right)$.}
In the same way, the other two diagrams of fig.\fig{Feyn} are given by
\begin{align}
&\left\langle\varphi_{\mathbf{k}_1} \varphi_{\mathbf{k}_{\mathbf{2}}} \varphi_{\mathbf{k}_{\mathbf{3}}} \right\rangle^{\prime}_{NNT}   =\frac{\pi^2\eta_N^2V_{NNT}H^{2}}{32i k_1^3k_3^3}\int_0^{\infty} \frac{d z}{z^4} \sum_{a= \pm} a\left(1+i a z\right) e^{-i a z}\mathcal{I}_a\left(\frac{k_1}{k_2} z\right) \mathcal{I}_a\left(\frac{k_3}{k_2} z\right)+
P,\label{B_b}\\
&\left\langle\varphi_{\mathbf{k}_1} \varphi_{\mathbf{k}_{\mathbf{2}}} \varphi_{\mathbf{k}_{\mathbf{3}}} \right\rangle^{\prime}_{NNN}=\frac{\pi^3\eta_N^3V_{NNN}H^{2}}{64 i k_2^3k_3^3}\int_0^{\infty} \frac{d z}{z^4} \sum_{a= \pm} a\  \mathcal{I}_a(z)\mathcal{I}_{a}\left(\frac{k_2}{k_1} z\right) \mathcal{I}_{a}\left(\frac{k_3}{k_1} z\right),\label{B_c}
\end{align}
where in eq.~\eqref{B_b}, the two permutations are $k_2 \leftrightarrow k_1$ and $k_2 \leftrightarrow k_3$.

\subsection{Squeezed limit}\label{SLimit}
The cosmological collider signal can be sharply seen in the squeezed limit $k_3\ll k_1\sim k_2$.\footnote{The full analytic expressions of bi-spectrum without taking the squeezed limit can be found in \cite{1811.00024, 2301.07047} for the single exchange diagram and in \cite{2404.09547} for the double exchange diagram.} 
In this limit, the first term of eq.~\eqref{B_a} 
dominate the other terms denoted as `permutations'. 
By using the approximation of eq.~\eqref{asymp} we obtain
\begin{align}
\left.\left\langle\varphi_{\mathbf{k}_1} \varphi_{\mathbf{k}_{\mathbf{2}}} \varphi_{\mathbf{k}_{\mathbf{3}}} \right\rangle^{\prime}_{NTT}\right|_{k_3\ll k_1\sim k_2} =
\frac{2\pi H^3 \eta_N  V_{NTT} C_{NTT}(\nu)}{k_1^3k_3^3 H}\left(\frac{k_3}{k_1}\right)^{3 / 2-\nu}+(\nu \rightarrow-\nu),
\end{align}
where the $z$-integral in eq.~\eqref{B_a} has been explicitly performed, obtaining the
$C_{NTT}$ of eq.\eq{CNTT}.
In the same way, the other diagrams are evaluated as  
\begin{align}
&\left.\left\langle\varphi_{\mathbf{k}_1} \varphi_{\mathbf{k}_{\mathbf{2}}} \varphi_{\mathbf{k}_{\mathbf{3}}} \right\rangle^{\prime}_{NNT}\right|_{k_3\ll k_1\sim k_2} =\frac{4\pi H^3\eta_N^2  V_{NNT} C_{NNT}(\nu)}{k_1^3k_3^3 H}\left(\frac{k_3}{k_1}\right)^{3 / 2-\nu}+(\nu \rightarrow-\nu),\\
&\left.\left\langle\varphi_{\mathbf{k}_1} \varphi_{\mathbf{k}_{\mathbf{2}}} \varphi_{\mathbf{k}_{\mathbf{3}}} \right\rangle^{\prime}_{NNN}\right|_{k_3\ll k_1\sim k_2} =\frac{2\pi H^3 \eta_N^3 V_{NNN} C_{NNN}(\nu)}{k_1^3k_3^3 H }\left(\frac{k_3}{k_1}\right)^{3 / 2-\nu}+(\nu \rightarrow-\nu),
\end{align}
obtaining the  $C_{NNT}$ and $C_{NNN}$ given in eq.s~(\ref{sys:Cs}).
The terms with $\nu \rightarrow-\nu$ are subdominant when $\nu$ is real, corresponding to $m_{\sigma}/H<3/2$.
Finally, as discussed in the text, the inflaton $\varphi$ is approximately converted into the curvature perturbation $\zeta$ as $\zeta=-H\varphi/\dot{\phi}_0$.


\end{appendix}

\end{document}